\documentclass[10pt,a4paper,pre,twocolumn,showpacs,superscriptaddress,bibnotes]{revtex4}
\usepackage{graphicx}
\usepackage{bm}
\usepackage{hyperref}
\usepackage{graphicx}
\usepackage{amsmath}
\usepackage{amssymb}
\usepackage{stmaryrd}
\usepackage{color}
\newcommand{\yaming}[1]{\textcolor{black}{#1}}

\begin{document}

\title{On large deviation properties of Brownian motion with dry friction}
\author{Yaming Chen}
\email{yaming.chen@qmul.ac.uk}
\author{Wolfram Just}
\email{w.just@qmul.ac.uk}

\affiliation{School of Mathematical Sciences, Queen Mary University of London, London E1 4NS, United Kingdom}

\date{September 17, 2014}

\begin{abstract}

We investigate piecewise-linear stochastic models as with regards to
the probability distribution of functionals of the stochastic processes,
a question which occurs frequently in large deviation
theory. The functionals that we are looking into in detail are
related to the time a stochastic process spends at a
phase space point or in a phase space region, as well as to the motion
with inertia. For a Langevin equation with discontinuous drift,
we extend the so-called backward Fokker-Planck technique for nonnegative
support functionals to arbitrary support functionals, to derive
explicit expressions for the moments of the functional.
Explicit solutions for the moments and for the distribution of
the so-called local time, the occupation time and the displacement
are derived for the Brownian motion with dry friction, including
quantitative measures to characterize deviation from Gaussian
behavior in the asymptotic long time limit.

\end{abstract}

\maketitle

\section{Introduction}

The study of functionals of stochastic processes
attracts considerable interest from various points of view,
most notably in the context of large deviation theory
where the probability distribution of integrals or sums of
random variables result in a potential characterizing large
deviations from the mean and extreme events
(see \cite{touchette2009} for a recent review).
Some simple functionals can characterize the time a process spends at a phase space point
or in a phase space region, or more simply such integrals may just represent
physical quantities like the position of an object \cite{Majumdar2005Brownian}.
Here we consider these functionals in the context of piecewise-smooth
stochastic systems. We use the motion of a Brownian particle subjected
to dry friction as an illustrative example where explicit expressions for the
distribution of functionals can be derived in a formally exact manner, and
where we can quantify the deviation from Gaussian behavior which prevails
in the asymptotic limit.

Functionals of a process $ v(t) $ have been investigated intensively in the past and have found numerous applications in physics \cite{Majumdar2005Brownian}. For instance, two of the popular functionals are the local time $ \int_0^t\delta(v(\tau))d\tau $ and the occupation time $ \int_0^t\theta(v(\tau))d\tau $ (see, e.g., \cite{SabMajumdar2006functionals}), which describe how much time the process $ v(t) $ has visited the origin and how long it has
taken positive values in the time window $ [0,t] $, respectively. Here $ \delta(v) $ is the Dirac delta function and $ \theta(v) $ is the step function with $ \theta(v)=1 $ for $ v>0 $ and $ \theta(v)=0 $ for $ v\leqslant 0 $. In addition, the area under the process, i.e., $ \int_0^t v(\tau) d\tau $, which is referred to as displacement if $ v(t) $ denotes the velocity of an object, is of particular importance due to its physical meaning.

To find statistical properties of functionals is usually nontrivial even for some of the simplest functionals, such as the local time and the occupation time. For the displacement, which is equivalent to solve a stochastic differential equation with inertial term, one often restricts the study to the overdamped case, in which limit the inertial term can be neglected \cite{Risken1989FPE}. To the best of our knowledge, the distributions of the displacement are only available in closed analytic form for the cases of the pure diffusion process and the Ornstein-Uhlenbeck process \cite{Kramers1940}. To compute the
distribution of a functional there are two significant achievements, i.e., the celebrated Feynman-Kac formula developed by Kac \cite{Kac1949} and the so-called backward Fokker-Planck technique proposed by Majumdar and Comtet \cite{MajumdarComtet2002OTD}.
The former enables one to obtain a corresponding Schr\"{o}dinger equation for the distribution of a functional and the later results in a backward Fokker-Planck equation for the distribution.
The advantage of the later over the former is that the later depends only on the initial condition of the process, no extra integral over $ v $ is required to obtain the distribution of the functional.

In this paper, we attempt to derive analytic results for some functionals of the Brownian motion with dry (also called solid or Coulomb) friction,
\begin{equation}\label{ad_ch2}
\dot{v}(t)=-\mu\sigma(v(t))-\gamma v(t)+\sqrt{D}\xi(t),
\end{equation}
which is used to describe the velocity of a solid object of unit mass subjected to a dry friction in addition to a viscous drag and a random noise (see, e.g., \cite{KawaradaHayakawa2004dry,Hayakawa2005Langevin,Gennes2005dryfriction,TouchetteStraeten2010Brownian}). Here
 $ \sigma(v) $, denoting the sign of $v$, represents the dry friction force with coefficient $ \mu>0 $, $ \gamma \geqslant 0 $ denotes the viscous coefficient and $ D>0 $ is the strength of the Gaussian white noise $ \xi(t) $
characterized by
\begin{equation}
\langle \xi(t)\rangle=0,\qquad \langle\xi(t)\xi(t')\rangle=2\delta(t-t').
\label{aa_ch2}
\end{equation}
The notation $ \langle\cdots\rangle $ stands for the average over all possible realizations of the noise.
For this piecewise-linear stochastic model, the transition probability distribution or propagator can be obtained by using spectral decomposition methods \cite{TouchetteStraeten2010Brownian} or Laplace transforms \cite{TouchetteThomas2012Brownian}. Especially, when $ \gamma=0 $, i.e.,  the model (\ref{ad_ch2}) with pure dry friction (also call Brownian motion with two-valued drift), the corresponding propagator is available in closed analytic form (see, e.g., \cite{karatzas1984,KaratzasShreve1991,TouchetteStraeten2010Brownian}). The weak-noise approximation of the model (\ref{ad_ch2}) has also been investigated in detail by using a path integral approach \cite{BauleCohenTouchette2010path,BauleTouchetteCohen2011path,ChenBaule2013}. In addition, the first-passage time distributions of Eq.~(\ref{ad_ch2}) can be derived by using spectral decomposition methods or solving the backward Kolmogorov equation (see \cite{ChenJust2014}). The dry friction effect in stochastic settings is also observed in experiments, such as those investigated recently by Chaudhury \textit{et al.} \cite{ChaudhuryMettu2008Brownian,GoohpattaderMettu2009Experimental,
GoohpattaderChaudhury2010Diffusive,GoohpattaderChaudhury2012} and Gnoli \textit{et al.} \cite{Gnoli2013BrownianRatchet,Gnoli2013GranularBrownian}.
Since the motion of two solid objects over each other is ubiquitous in nature \cite{LiDong2011PRL,baule2012,BauleSollich2013,Weymouth2013PRL,SanoHayakawa2014}, the dry friction model (\ref{ad_ch2}) plays a significant role at the interface between theory and experiment. Therefore, it is interesting to have a close look at stochastic properties of functionals whose underlying dynamics is governed by this model. In particular we will focus on the local time, the occupation time and the displacement in this paper.

It is worth noting that the local time and the occupation time of the pure dry friction case [i.e., $ \gamma=0 $ in Eq.~(\ref{ad_ch2})] have been investigated in \cite{MajumdarComtet2002OTD,SabMajumdar2006functionals} even though the authors did not attempt to study the dry friction effect in a stochastic setting.
An integral representation of the occupation time distribution of this simple model has also been derived in \cite{Simpson2013opt}. But the recursive relations for the moments of the local time and the occupation time have not been
given explicitly. In addition, to the best of our knowledge analytic results of the displacement for the pure dry friction case and functionals of the full model (\ref{ad_ch2}) are not available in the literature.

The rest of this paper is arranged as follows. We first extend in Sec.~\ref{sec2} the backward Fokker-Planck technique for a positive support functional to an arbitrary support functional by replacing the double Laplace transform used in \cite{MajumdarComtet2002OTD} with a Fourier-Laplace transform. Rather than solving the backward equation directly, we then derive from this equation a recursive ordinary differential equation (ODE) for the moments of the functional.
In Secs.~\ref{sec3} and \ref{sec4}, we show that the moments of the local time and the occupation time are given explicitly by the solution of the corresponding homogeneous ODE. Then we apply the results to solve the dry friction model (\ref{ad_ch2}).
In Sec.~\ref{sec5}, even though a simple formula is not available for a general drift, we show that the moments of the displacement can be obtained in the Laplace space for the dry friction model (\ref{ad_ch2}). Finally, results are summarized in Sec.~\ref{sec6}.

\section{Remarks on general piecewise-smooth stochastic systems}
\label{sec2}

Let us consider the functional
\begin{equation}\label{mya}
T=\int_0^tU(v(\tau))d\tau
\end{equation}
for an integrable kernel $ U(v) $. The stochastic process
$ v(t) $ with initial condition $v(0)=v_0$
is assumed to obey the following Langevin equation
\begin{equation}\label{ca_ch2}
\dot{v}(t)=-\Phi'(v(t))+\xi(t)
\end{equation}
with potential $ \Phi(v) $, where $ \xi(t) $ denotes the Gaussian white noise defined by Eq.~(\ref{aa_ch2}).
In what follows, we will first discuss some general properties of the distribution $ p(T,t,v_0) $
of the functional (\ref{mya}). In the latter parts we are then going to derive
relations for the moments $ M_n(t,v_0) $ of $T$ which will be used in the applications.

\subsection{Distribution of the functional}
\label{subii1}

While it may be difficult to derive closed analytic expressions for the distribution $ p(T,t,v_0) $
of the functional (\ref{mya}),  some statements can be made on the asymptotic properties in the limit of large $t$
if we assume that the potential of the Langevin equation (\ref{ca_ch2}) is stable, i.e., $ \Phi(v)\rightarrow \infty $
when $ v\rightarrow \pm \infty $, and that the correlation of the corresponding process decays sufficiently fast. Then a simple heuristic argument suggests (see, e.g., \cite{SabMajumdar2006functionals}) that
the distribution $ p(T,t,v_0) $ is simply Gaussian around its mean in the long time limit.
Thus the limiting distribution is characterized only by the first and the second moments,
\begin{equation}\label{ec_ch5}
p^{\text{asym}}(T,t,v_0)=\frac{1}{\sqrt{2\pi \mathrm{Var}(T)}}\exp\left[-\frac{(T-M_1(t,v_0))^2}{2\mathrm{Var}(T)}\right],
\end{equation}
where $ M_1(t,v_0) $ and $ \mathrm{Var}(T) $ stand for the first moment and the variance of $ T $, respectively.
As stated in \cite{SabMajumdar2006functionals}, the argument is as follows: At large time $ t $, assuming the process
$ v(t) $  to be mixing, the corresponding propagator $ p(v,t|v_0,0) $ tends to the stationary distribution
$ \exp[-\Phi(v)]/Z $, where $ Z $ is a normalization factor. Hence,  at time increments which exceed
the correlation time of the system, the random variables $ U(v(\tau))-\langle U(v(\tau))\rangle $ are only weakly
correlated. Therefore, from the definition (\ref{mya})
we expect a central limit theorem to hold in the limit of $ t $ being much larger than the relaxation time
and the correlation time of the two aforementioned variables, resulting finally in a Gaussian distribution for the
functional $ T $ around its mean. Hence, in order to determine the limiting distribution of $ T $
around its mean we only need to obtain the first and the second moments, which can be attained by using the results
stated in the following subsections. Our arguments will be able to predict higher-order cumulants as well
so that we provide a quantitative tool to benchmark the aforementioned reasoning.

For the subsequent investigations we will need the boundary conditions of the distribution $ p(T,t,v_0) $
in the limit when $ v_0 $ tends to infinity. These boundary conditions depend on the particular
function $ U(v) $ which enters the functional $T$. They can be derived from the observation
that a process starting at infinity cannot cross the origin in finite time. For simplicity let us discuss
two particular choices, namely $U(v)=\delta(v)$ and $U(v)=\theta(v)$, corresponding to the so-called
local time
\begin{equation}\label{myd}
T_{\text{loc}}= \int_0^t \delta(v(\tau)) d\tau
\end{equation}
which measures the time that the process visits the origin, and the so-called occupation time
\begin{equation}\label{myg}
T_{\text{occ}} =\int_0^t \theta(v(\tau)) d \tau
\end{equation}
which measures the time that the process takes positive values.
Hence for the distribution of the local time we have
\begin{equation}
\label{bi_ch5}
 p_{\text{loc}}(T,t,v_0\rightarrow \pm \infty)=\delta(T),
\end{equation}
and for the occupation time the boundary condition reads
\begin{eqnarray}
&& p_{\text{occ}}(T,t,v_0\rightarrow \infty)=\delta(T-t),\nonumber\\
&& p_{\text{occ}}(T,t,v_0\rightarrow -\infty)=\delta(T). \label{bk_ch5}
\end{eqnarray}
The third quantity we are going to discuss in some detail will be the displacement of the particle which
can also be written as a functional with the choice $U(v)=v$, i.e.,
\begin{equation}
T_{\text{dis}}=\int_0^t v(\tau) d \tau.
\end{equation}
For the displacement, the argument of the boundary condition is slightly more subtle. We expect that for $ |v_0|\gg t $ the
velocity of the particle does not change considerably in the time window $ [0,t] $ and that the displacement should be at
the scale of $ v_0 t $. Therefore, the corresponding boundary condition can be formally approximated as
$p_{\text{dis}}(T,t,v_0)\sim \delta(T-v_0 t)$
in the asymptotic limit $ v_0 \rightarrow \pm\infty $. Such a condition will be used later to obtain the boundary condition for the moments of the displacement.

\subsection{Moments of the functional}
We start to derive one of the main results by assuming the potential $ \Phi(v) $ to be continuous and to be smooth
everywhere apart from $ v=0 $ where we allow the potential to be not differentiable, so that the deterministic part
of the Langevin equation (\ref{ca_ch2}) has a discontinuity at $v=0$. More general cases of
piecewise-smooth systems could be considered with ease but the discussion of this special case will be sufficient
to uncover the general structure. In addition, our particular choice also contains the dry friction model (\ref{ad_ch2}).
For a nonnegative support functional it has been shown by Majumdar and Comtet \cite{MajumdarComtet2002OTD} that the
double Laplace transform of the distribution $ p(T,t,v_0) $ satisfies a backward equation \yaming{(see Eq.~(4) in  \cite{MajumdarComtet2002OTD})}. \yaming{In this paper we do not confine the discussion to nonnegative functionals, i.e., $ T $ can be negative. Hence, instead of using the double Laplace transform we have to consider the Fourier-Laplace transform of the distribution $ p(T,t,v_0) $, i.e.,}
\begin{equation}
r(k,s,v_0)=\int_{-\infty}^{\infty}e^{-i k T} dT \int_0^{\infty} e^{-s t} p(T,t,v_0)dt.
\label{ad}
\end{equation}
Similarly, using the method stated in \cite{MajumdarComtet2002OTD} (see also \cite{SabMajumdar2006functionals})
it is straightforward to show that $ r(k,s,v_0) $ satisfies the following backward equation
\begin{widetext}
\begin{eqnarray}
\partial^2_{v_0} r(k,s,v_0)-\Phi'(v_0)\partial_{v_0}r(k,s,v_0)-[ikU(v_0)+s] r(k,s,v_0)+1=0. \label{ae}
\end{eqnarray}
Let us assume that the $n$-th moment $ M_n(t,v_0) $ of $ T $ exists so that the moments of $ T $ in the
Laplace space are given by
\begin{eqnarray}
\widetilde{M}_{n}(s,v_0)=\int_0^\infty e^{-st }M_n(t,v_0)dt=\left. i^n \partial^n_{k}r(k,s,v_0)\right |_{k=0}
&& \mbox{for } n \geqslant 1.
\end{eqnarray}
Thus by acting the operator $ i^n\partial^n_k $ on Eq.~(\ref{ae}) and setting $ k=0 $,
we obtain a hierarchy of equations for $ \widetilde{M}_{n}(s,v_0) $,
\begin{eqnarray}
\partial^2_{v_0} \widetilde{M}_{n}(s,v_0)-\Phi'(v_0)\partial_{v_0} \widetilde{M}_{n}(s,v_0)-s \widetilde{M}_{n}(s,v_0)
=-nU(v_0)\widetilde{M}_{n-1}(s,v_0),\qquad \widetilde{M}_0(s,v_0)=r(0,s,v_0)=1/s,
\label{bd_ch5}
\end{eqnarray}
\end{widetext}
which is easier to deal with than solving Eq.~(\ref{ae}).

\subsection{Boundary conditions}
\label{subsec1}
In order to solve Eq.~(\ref{bd_ch5}), it is essential to know the boundary conditions for the moments in the
limit $ v_0\rightarrow \pm \infty$. Using Eq.~(\ref{bi_ch5}) and the uniform convergence of the Laplace transform
for $\mbox{Re}(s)>0$ it follows that the moments of the local time obey
\begin{equation}
\widetilde{M}^{\text{loc}}_n(s,v_0\rightarrow \pm \infty)=0 \qquad \mbox{for } \mbox{Re}(s)>0 \mbox{ and } n\geqslant 1,
\label{bj_ch5}
\end{equation}
whereas a similar argument and Eq.~(\ref{bk_ch5}) yield for the moments of the occupation time that
\begin{equation}
\widetilde{M}^{\text{occ}}_n(s,v_0\rightarrow \infty)=\frac{n!}{s^{n+1}},\qquad
 \widetilde{M}^{\text{occ}}_n(s,v_0\rightarrow -\infty)=0
 \label{bm_ch5}
\end{equation}
for $ \mbox{Re}(s)>0 $  and $  n\geqslant 1 $.
For the displacement, the reasoning presented at the end of subsection
\ref{subii1} suggests that the $ n$-th moment of the displacement as a function of $ v_0 $ should be bounded by a polynomial of order $n$ in $ v_0 $. We will see later that this condition is sufficient
to determine the solution of the dry friction model (\ref{ad_ch2}).

\subsection{Matching conditions}
\label{subsec2}
Since we are here concerned with a piecewise-smooth potential $ \Phi(v) $, say with a kink at $v=0$,
we have to solve Eq.~(\ref{bd_ch5}) for $ v_0>0 $ and $ v_0<0 $, respectively.
Even for a piecewise-smooth potential $\Phi(v)$ the solution of Eq.~(\ref{bd_ch5})
has to be at least continuous, so that we require continuity of the moments
at $ v_0=0 $, resulting in the first matching condition
\begin{equation}
\label{bo_ch5}
\widetilde{M}_n(s,0-)=\widetilde{M}_n(s,0+) .
\end{equation}
As for the first derivative we may obtain a matching condition by integrating
Eq.~(\ref{bd_ch5}) across the discontinuity at $v_0=0$. Because of the continuity of the moments the right hand side will contribute only if the kernel $U(v)$ has a $\delta$ singularity at
$v=0$. Thus for the local time (\ref{myd}) the matching condition reads
\begin{equation}
\partial_{v_0} \widetilde{M}^{\text{loc}}_n(s,0+)-\partial_{v_0} \widetilde{M}^{\text{loc}}_n(s,0-)
=-n \widetilde{M}^{\text{loc}}_{n-1}(s,0),
\label{bj0_ch5}
\end{equation}
whereas for any other kernel, e.g., for the occupation time and for the displacement the matching condition just
simplifies to
\begin{equation}
\label{bq_ch5}
\partial_{v_0} \widetilde{M}^{\text{occ}/\text{dis}}_n(s,0+)=\partial_{v_0}
\widetilde{M}^{\text{occ}/\text{dis}}_n(s,0-) .
\end{equation}
Here we have used the shorthand notation $ \partial_{v_0}\widetilde{M}_n(s,0\pm) $ to
denote $  \partial_{v_0}\widetilde{M}_n(s,v_0)|_{v_0\rightarrow 0\pm} $.

\subsection{Structure of the solution}

Let us briefly discuss how we are going to approach the analytic solution of the hierarchy
(\ref{bd_ch5}) for a general potential $\Phi(v)$ with nonanalyticity at $v=0$. Suppose
that the appropriate fundamental piecewise-smooth solution $ \varphi(s,v_0) $
of the corresponding homogeneous ODE of Eq.~(\ref{bd_ch5}) is known, which vanishes for $ \mbox{Re}(s)>0 $
when $ v_0\rightarrow \pm \infty $. The solution obeys
\begin{equation}
\left[\partial_{v_0}^2-\Phi'(v_0)\partial_{v_0}-s \right] \varphi(s,v_0)=0
\label{by_ch5}
\end{equation}
for $v_0\neq 0$ and we assume $ \varphi(s,v_0) $ to be continuous at $ v_0=0 $.
To start with, we consider the hierarchy (\ref{bd_ch5}) separately on the two domains $v_0>0$ and
$v_0<0$. Given $\widetilde{M}_0(s,v_0)=1/s$ [see Eq.~(\ref{bd_ch5})] we denote by $\widetilde{M}^p_1(s,v_0)$ a particular solution
of the inhomogeneous differential equation (for $n=1$) which fulfils the boundary condition at
infinity. By adding a multiple of the homogeneous solution $ \varphi(s,v_0) $ we then obtain
two branches with two constants of integration which are adjusted according to the
matching conditions (see subsection \ref{subsec2}). Hence, at each level of
the hierarchy we determine a particular solution $\widetilde{M}^p_n(s,v_0)$ and construct the general
solution, which already satisfies the boundary condition at infinity (see subsection \ref{subsec1}), through
\begin{equation}\label{bh0_ch5}
\widetilde{M}_n(s,v_0)=\widetilde{M}_n^p(s,v_0)+\left\{
\begin{array}{ll}
C^+_n\varphi(s,v_0), & v_0>0,\\
C^-_n\varphi(s,v_0), & v_0<0 .
\end{array}
\right.
\end{equation}
Finally we compute the coefficients $ C^{\pm}_n $ with the help of the matching conditions at $ v_0=0 $
(see subsection \ref{subsec2}).
The relevance of the trivial Eq.~(\ref{bh0_ch5}) comes from the observation that
we just have to deal with the matching conditions caused by the nonanalyticities of the potential.
Furthermore, in most of the cases considered here, we will be able to express the particular solutions
$\widetilde{M}^p_n(s,v_0)$ in terms of the homogeneous solution $\varphi(s,v_0)$ as well, so that the entire
structure just requires solving Eq.~(\ref{by_ch5}).

We may not be able to do the inverse Laplace transform for the moments analytically. However, in the long time limit
the behavior of the moments is dominated by the singular terms of the Laplace transform
and those are often not too difficult to evaluate. We can also resort to numerical Laplace inversion, such as
the so-called Talbot method \cite{AbateValkoLaplace2004}, which usually gives accurate results even at a very short time.

In the next three sections, we will apply the formula (\ref{bh0_ch5}) to
the local time, the occupation time, and the displacement. The analysis of the local time
can be largely done for general potential while we are going to derive explicit formulas for
the other two observables for the dry friction model (\ref{ad_ch2}) without and with viscous friction, respectively.

\section{Local time}
\label{sec3}
To begin with we first consider the local time problem, i.e., the setup with kernel $U(v)=\delta(v)$. In this case
it is possible to derive closed form expressions for a general potential $\Phi(v)$ with discontinuity at $v=0$.
We will use this simple case to illustrate our approach.

\subsection{General drift}
As the matching condition (\ref{bj0_ch5}) takes care of the singularities at $v=0$ we only need to
consider the hierarchy (\ref{bd_ch5}) for nonvanishing arguments, i.e., we can confine our study to the homogeneous equation
(\ref{by_ch5}). Hence in Eq.~(\ref{bh0_ch5}) the particular solution vanishes and the moments are expressed
in terms of the fundamental solution $\varphi(s,v_0)$. The coefficients follow immediately
from the matching conditions at $v_0=0$ [see Eqs.~(\ref{bo_ch5}) and  (\ref{bj0_ch5})] and we arrive at
the recursive relation
\begin{equation}
\widetilde{M}^{\text{loc}}_n(s,v_0)=
\frac{n\varphi(s,v_0)\widetilde{M}^{\text{loc}}_{n-1}(s,0)}{\partial_{v_0}\varphi(s,0-)
-\partial_{v_0}\varphi(s,0+)} \qquad \mbox{for } n\geqslant 1. \label{bf_ch5}
\end{equation}
Therefore, the higher moments are determined by the first moment via
\begin{equation}\label{bg_ch5}
\widetilde{M}^{\text{loc}}_n(s,v_0)=n!
\big[s\widetilde{M}^{\text{loc}}_1(s,0)\big]^{n-1}
\widetilde{M}^{\text{loc}}_1(s,v_0),
\end{equation}
where
\begin{equation}
\widetilde{M}^{\text{loc}}_1(s,v_0)=
\frac{\varphi(s,v_0)/s}{\partial_{v_0}\varphi(s,0-)
-\partial_{v_0}\varphi(s,0+)}.
\label{bg1_ch5}
\end{equation}
This simple relation enables one to obtain the distribution of the local time explicitly in the Fourier-Laplace space, which reads [see Eq.~(\ref{ad})]
\begin{eqnarray}
r(k,s,v_0)&=&\sum_{n=0}^{\infty}\frac{(-i k)^n}{n!}\widetilde{M}^{\text{loc}}_n(s,v_0)\nonumber\\
&=&
\frac{\widetilde{M}^{\text{loc}}_1(s,v_0)}{s\widetilde{M}^{\text{loc}}_1(s,0)}\frac{1}{1+ik  s
\widetilde{M}^{\text{loc}}_1(s,0)}.
\end{eqnarray}
Then inverting the Fourier transform with respect to $ k $ yields the Laplace transform of the
distribution for the local time
\begin{equation}
\widetilde{p}_{\text{loc}}(T,s,v_0)
=\frac{\widetilde{M}^{\text{loc}}_1(s,v_0)}
{s^2[\widetilde{M}^{\text{loc}}_1(s,0)]^2}\exp\left(-\frac{T}{s\widetilde{M}^{\text{loc}}_1(s,0)}\right)
\label{bg0_ch5}
\end{equation}
for $  T\geqslant 0 $, which indicates that the distribution is fully determined by the Laplace transform of
its first moment. The result (\ref{bg0_ch5}) extends that obtained in \cite{SabMajumdar2006functionals} to be valid
for arbitrary initial condition $ v_0 $ [see Eq.~(30) therein]. In general, we may not be able to do the inverse
Laplace transform for Eq.~(\ref{bg0_ch5}) analytically. However, we can resort to numerical inversion
(e.g., the Talbot method \cite{AbateValkoLaplace2004}) to produce the distribution
$ p_{\text{loc}}(T,t,v_0) $ numerically, especially for short time $t$.

\subsection{Pure dry friction}
In order to make further progress and to illustrate the effectiveness of the approach
let us consider the particular model (\ref{ad_ch2}) with only dry friction ($\gamma=0$).
In this case, we let $ \mu=D=1 $. Other values can be covered by using the appropriate rescaling
\begin{equation}
x=\mu v/D, \qquad \tau=\mu^2 t/D.
\end{equation}
Hence the pure dry friction case can be written in the form (\ref{ca_ch2})
with a stable potential
\begin{equation}
\Phi(v)=|v|.
\label{bh_ch5}
\end{equation}
The appropriate fundamental solution of the homogeneous ODE (\ref{by_ch5}) corresponding to Eq.~(\ref{bd_ch5}) is
\begin{equation}
\varphi(s,v_0) =e^{(1-\sqrt{1+4s})|v_0|/2},
\label{da_ch5}
\end{equation}
which vanishes for $ \mbox{Re}(s)>0 $ when $ v_0\rightarrow \pm \infty $.
Then, from Eqs.~(\ref{bg0_ch5}) and (\ref{bg1_ch5}) we obtain
the local time distribution in the Laplace domain,
\begin{equation}
\widetilde{p}_{\text{loc}}(T,s,v_0)=\frac{\sqrt{1+4s}-1}{s}e^{ -(\sqrt{1+4s}-1)(T+|v_0|/2) }
\label{dg0_ch5}
\end{equation}
for $ T\geqslant 0 $. In this case we can even perform the inverse Laplace transform explicitly.
From the table of Laplace transforms in \cite{AbramovitzStegun1968} we have the identity
\begin{equation}
\mathcal{L}^{-1}\left(\frac{e^{-k\sqrt{\alpha}}}{b+\sqrt{\alpha}}\right)=\frac{e^{-k^2/(4t)}}{\sqrt{\pi t}}-be^{bk+b^2t}\mathrm{erfc}\left(b\sqrt{t}+\frac{k}{2\sqrt{t}}\right),
\end{equation}
where $ \mathrm{erfc}(z) $ is the complementary error function.
Hence, using the shifting property of Laplace transform, we obtain from Eq.~(\ref{dg0_ch5}) the inverse Laplace transform,
i.e, the local time distribution
\begin{eqnarray}
\hspace{-2em} p_{\text{loc}}(T,t,v_0)
 &=&
\frac{2}{\sqrt{\pi t}}e^{-(2T+|v_0|-t)^2/(4t)}\nonumber\\
&&-e^{ 2T+|v_0| }\mathrm{erfc}\left(\frac{\sqrt{t}}{2}
+\frac{2T+|v_0|}{2\sqrt{t}}\right)
\label{bg2_ch5}
\end{eqnarray}
for $ T\geqslant 0 $, which generalizes the result in \cite{SabMajumdar2006functionals} to be valid for any $ v_0 $ [see Eq.~(A4) therein].

\begin{figure*}
\begin{center}
\includegraphics[scale=0.9]{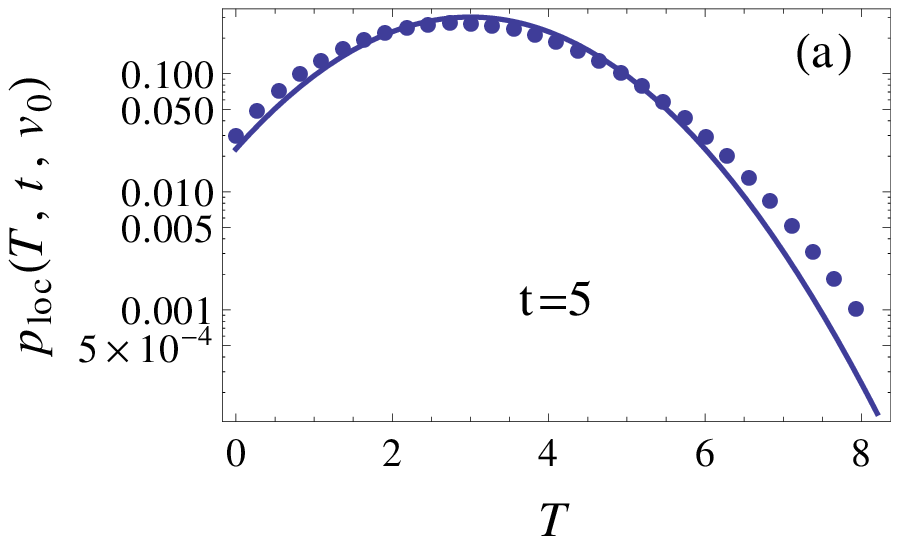}
\includegraphics[scale=0.9]{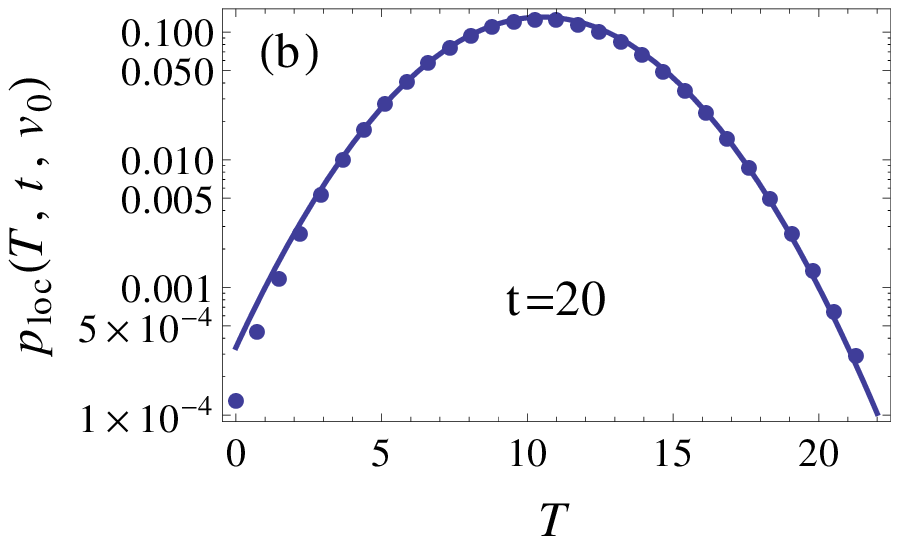}
\end{center}
\caption{(Color online) Local time distribution of the pure dry friction case [see Eqs.~(\ref{ca_ch2}) and (\ref{bh_ch5})] for $ v_0=0 $  and two different times: (a) $ t=5 $ and (b) $ t=20 $. Lines correspond to the leading order limiting distribution (\ref{df1_ch5}), and points to the analytic expression (\ref{bg2_ch5}). Deviations from
Gaussian behavior are noticeable for small time only.}
\label{fig_loc_puredry}
\end{figure*}

In most of cases the inverse Laplace transform cannot be performed analytically. Thus let us discuss
how the previous analysis based on moments [see, e.g., Eq.~(\ref{ec_ch5})] matches with the exact
expression (\ref{bg2_ch5}). We will also use the opportunity to illustrate the method to derive
asymptotic results, which will be used later in cases where no exact distribution function is available.
Using  Eqs.~(\ref{bg_ch5}), (\ref{bg1_ch5}) and (\ref{da_ch5}) the Laplace transform of the moments of the local
time are explicitly given
by
\begin{equation}\label{dc_ch5}
\widetilde{M}^{\text{loc}}_n(s,v_0) =\frac{n!}{4^ns^{n+1}}(\sqrt{1+4s}+1)^ne^{(1-\sqrt{1+4s})|v_0|/2} .
\end{equation}
The expressions have a pole at $ s=0 $ and a branch cut on the real axis for $ s<-1/4 $.
As for the inverse transform we can use the standard asymptotic result (see, e.g., \cite{Murray1974Asymptotics}) that the
time dependence of $M_n^{\text{loc}}(t,v_0)$ is determined by the pole at $s=0$ up to
contributions of $O(e^{-(1/4-o)t})$. \yaming{Here the notation $ o $ stands for an arbitrary small positive correction to $ 1/4 $. This correction is due to a power-law correction to the leading exponential behavior. The same notation will be used in the rest of this paper.} Hence the moments in the asymptotic long time
limit read
\begin{eqnarray}
&& \hspace{-2em} M^{\text{loc}}_n(t,v_0)\nonumber\\
&=&4^{-n}\left. \frac{d^n}{d s^n}\left[(\sqrt{1+4s}+1)^ne^{(1-\sqrt{1+4s})|v_0|/2+s t}\right] \right|_{s=0}\nonumber\\
&&+
O(e^{-(1/4-o)t}).
\label{db0_ch5}
\end{eqnarray}
In particular, for the first two moments we have the simple expressions
\begin{eqnarray}
M^{\text{loc}}_1(t,v_0)&=&\frac{t}{2}+\frac{1-|v_0|}{2}+O(e^{-(1/4-o)t}), \label{dd_ch5}
\end{eqnarray}
\begin{eqnarray}
M^{\text{loc}}_2(t,v_0) &=&  \frac{t^2}{4}+\frac{(2-|v_0|)t}{2}+\frac{v_0^2-2|v_0|-2}{4}\nonumber\\
&&+O(e^{-(1/4-o)t}),\label{de_ch5}
\end{eqnarray}
which result in the variance
\begin{equation}
\text{Var}(T_{\text{loc}})=\frac{t}{2}-\frac{3}{4}+O(e^{-(1/4-o)t}) .
\end{equation}
Finally  the limiting distribution (\ref{ec_ch5}) reads
\begin{eqnarray}
&& \hspace{-4em }  p_{\text{loc}}^{\text{asym}}(T,t,v_0)\nonumber\\
 &=& \frac{1}{\sqrt{\pi (t-3/2)}} \exp\left[ -\frac{(2T-t+|v_0|-1)^2}{4t-6} \right]\nonumber\\
 &&+O(e^{-(1/4-o)t}).
 \label{df1_ch5}
\end{eqnarray}
As seen in Fig.~\ref{fig_loc_puredry}, this limiting distribution  in leading order matches well
with the exact analytic expression (\ref{bg2_ch5}) at large time $ t $.

From the perspective of large deviation results it is of interest to look at the
asymptotic properties of the higher order
cumulamnts as well. Of course they can be worked out rather straightforwardly from the expression
(\ref{db0_ch5}). For instance, for the third and the fourth cumulants we obtain
\begin{eqnarray}
&&K^{\text{loc}}_3=7/4+O(e^{-(1/4-o)t}), \nonumber\\  &&K^{\text{loc}}_4=-45/8+O(e^{-(1/4-o)t}),
\label{dfa_ch5}
\end{eqnarray}
a result which shows that both are negligible compared to the first two moments
[see Eqs.~(\ref{dd_ch5}) and (\ref{de_ch5})] in the asymptotic limit of large $t$.
Hence this simple indicator confirms that the large deviation
function of the local time, defined by $p_{\text{loc}}(t \vartheta,t,v_0) \sim \exp(-t \phi(\vartheta))$
 is quadratic (see also Fig.~\ref{fig_loc_puredry}),  a feature which of course can be confirmed as well
if one employs the exact analytic expression of the distribution function (\ref{bg2_ch5}) and evaluates $t^{-1}\ln p_{\text{loc}}(t \vartheta, t, v_0)$
in the limit $t\rightarrow \infty$ (see also \cite{SabMajumdar2006functionals}).

\subsection{Dry and viscous friction}
\label{subsec_cc}

Let us now add viscous friction to the setup discussed in the previous subsection. Intuitively one probably would not
expect any major change as a linear viscous force is unlikely to have any impact on the predominantly Gaussian
behavior found previously.
For the full force case (\ref{ad_ch2}), we let $ \gamma=D=1 $ without loss of generality.
Other values are covered by the rescaling
\begin{equation}
 x=(\gamma/D)^{1/2}v,\qquad \tau=\gamma t.
\end{equation}
Hence the full model can be written in the form (\ref{ca_ch2}) with a stable potential
\begin{equation}\label{ea_ch5}
\Phi(v)=(|v|+\mu)^2/2.
\end{equation}
The appropriate fundamental solution of the homogeneous ODE (\ref{by_ch5}) is given by
\begin{eqnarray}
\varphi(s,v_0)=e^{(|v_0|+\mu)^2/4}D_{-s}(|v_0|+\mu), \label{eb_ch5}
\end{eqnarray}
which vanishes for $ \mbox{Re}(s)>0 $ in the limit $ v_0\rightarrow \pm \infty $.
Here we use the symbol $ D_{\nu}(z) $ to denote the parabolic cylinder functions \cite{Buchholz1969Hypergeometric}.

\begin{figure}
\begin{center}
\includegraphics[scale=0.9]{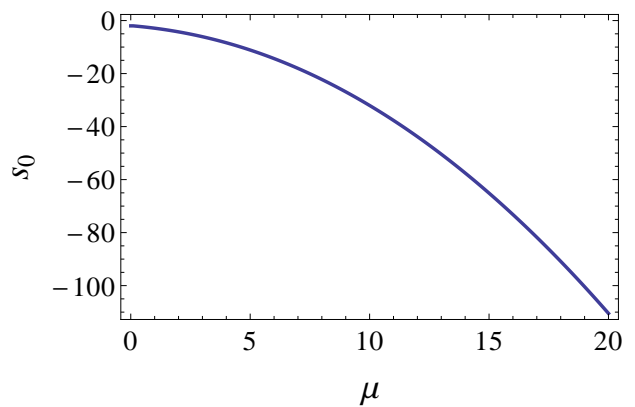}
\end{center}
\caption{(Color online)
Maximal solution $ s=s_0 $ (\ref{ed1_ch5}) of the equation $ D_{-s-1}(\mu)=0 $ as a function of $ \mu $.}
\label{fig_s0}
\end{figure}

\begin{figure*}
\begin{center}
\includegraphics[scale=0.9]{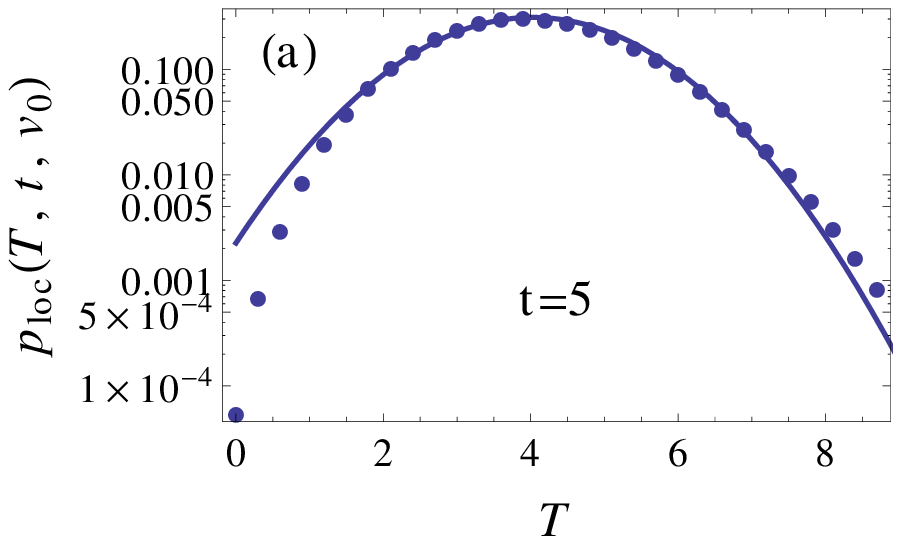}
\includegraphics[scale=0.9]{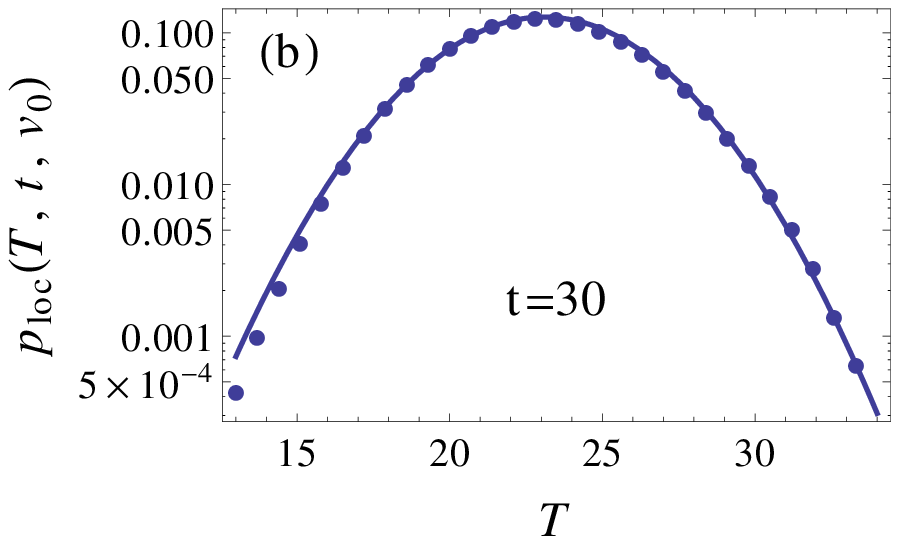}
\end{center}
\caption{(Color online)
Local time distribution of the dry and viscous friction case [see Eqs.~(\ref{ca_ch2}) and (\ref{ea_ch5})] for $ v_0=0 $, $ \mu=1 $ and two different times: (a) $ t=5 $ and (b) $ t=30 $. Lines corresponds to the leading-order limiting distribution (\ref{ec_ch5}),  and points to the numerical evaluation of Eq.~(\ref{bg0_ch5}).}
\label{fig_loc_dryviscous}
\end{figure*}

As already stated in the previous subsection we now employ the general expression (\ref{bh0_ch5})
without any particular inhomogeneous solution and use the matching conditions (\ref{bo_ch5}) and (\ref{bj0_ch5}) to compute
the Laplace transform of the moments. In fact the general expressions
Eqs.~(\ref{bg_ch5}) and (\ref{bg1_ch5}) immediately yield
\begin{eqnarray}
\widetilde{M}^{\text{loc}}_n(s,v_0)&=&\frac{n!}{2^n s^{n+1}} \left( \frac{D_{-s}(\mu)}{D_{-s-1}(\mu)}\right)^{n-1}\nonumber\\
&& \times \frac{e^{(|v_0|+\mu)^2/4} D_{-s}(|v_0|+\mu)}{ e^{\mu^2/4}D_{-s-1}(\mu)}. \label{ed_ch5}
\end{eqnarray}
For real argument $ z $ the parabolic cylinder function fulfils $ D_{\nu}(z)\neq 0$ for any $ \nu $ with nonvanishing imaginary part and $ D_{\nu}(z)> 0$ for any negative real value of $ \nu $. Hence, we conclude that all the singularities of Eq.~(\ref{ed_ch5}) lie on the
nonpositive real axis as $ \mu>0 $.
In addition all these singularities are poles. The largest negative pole which determines
the asymptotic properties of the moments is given by
\begin{equation}
s_0 =\max\{ s: D_{-s-1}(\mu)=0 \}<-1.
\label{ed1_ch5}
\end{equation}
The dependence of this value on the dry friction coefficient $\mu$ is displayed in  Fig.~\ref{fig_s0} and turns out
to be a monotonically decreasing function. It is now again straightforward to write down the leading order asymptotic expansion
of the inverse Laplace transform, i.e., of the time dependent moment, resulting in
\begin{widetext}
\begin{eqnarray}
M^{\text{loc}}_n(t,v_0)  =\left. \frac{ e^{(|v_0|+\mu)^2/4} }{2^{n}e^{\mu^2/4}}   \frac{d^n}{d s^n}  \left[\left( \frac{D_{-s}(\mu)}{D_{-s-1}(\mu)}\right)^{n-1} \frac{ D_{-s}(|v_0|+\mu)}{ D_{-s-1}(\mu)} e^{s t}\right] \right |_{s=0}+
O(e^{(s_0+o) t}), \label{ed0_ch5}
\end{eqnarray}
\yaming{where $ s_0+o<0 $. }
In particular the leading terms of the the first two moments in the long time limit are given explicitly as follows,
\begin{gather}
M^{\text{loc}}_1(t,v_0) =
\frac{D_0(\mu)t}{2D_{-1}(\mu)}+\frac{D_0(\mu)}{2D_{-1}(\mu)}\left( \frac{D_{-1}^{(1,0)}(\mu)}{D_{-1}(\mu)}
-\frac{D_{0}^{(1,0)}(\mu+|v_0|)}{D_{0}(\mu+|v_0|)}\right)+O(e^{(s_0+o) t}),\\
M^{\text{loc}}_2(t,v_0)=  \frac{D_0^2(\mu)t^2}{4D_{-1}^2(\mu)}+ \frac{D_0^2(\mu)}{2 D_{-1}^2(\mu)}\left(
2\frac{D_{-1}^{(1,0)}(\mu)}{D_{-1}(\mu)}
-\frac{D_{0}^{(1,0)}(\mu)}{D_{0}(\mu)}-\frac{D_{0}^{(1,0)}(\mu+|v_0|)}{D_{0}(\mu+|v_0|)}
\right) t+O(1).
\end{gather}
\end{widetext}
Here the symbol $ D^{(1,0)}_{\Lambda}(z)$ denotes the derivative of the parabolic cylinder function with
respect to its index, i.e., $\left.\partial_{\nu}D_{\nu}(z)\right|_{\nu=\Lambda} $.
Thus we obtain for the leading order term of the variance the expression
\begin{equation}\label{eh_ch5}
\mathrm{Var}(T_{\text{loc}})= \frac{D_0^2(\mu)}{2 D_{-1}^2(\mu)}\left(
\frac{D_{-1}^{(1,0)}(\mu)}{D_{-1}(\mu)}-\frac{D_{0}^{(1,0)}(\mu)}{D_{0}(\mu)}
\right) t+O(1).
\end{equation}
Here we have stated the results for the second moment and the variance up to $ O(1) $ as the constant term is too cumbersome to be written down explicitly. Of course, including the constant term the expression is again correct up to $ O(e^{(s_0+o) t}) $.
As shown in Fig.~\ref{fig_loc_dryviscous}, the corresponding asymptotic distribution (\ref{ec_ch5}) matches well with the numerical evaluation of
Eq.~(\ref{bg0_ch5}) for times which are sufficiently large.

\begin{figure*}
\begin{center}
\includegraphics[scale=0.9]{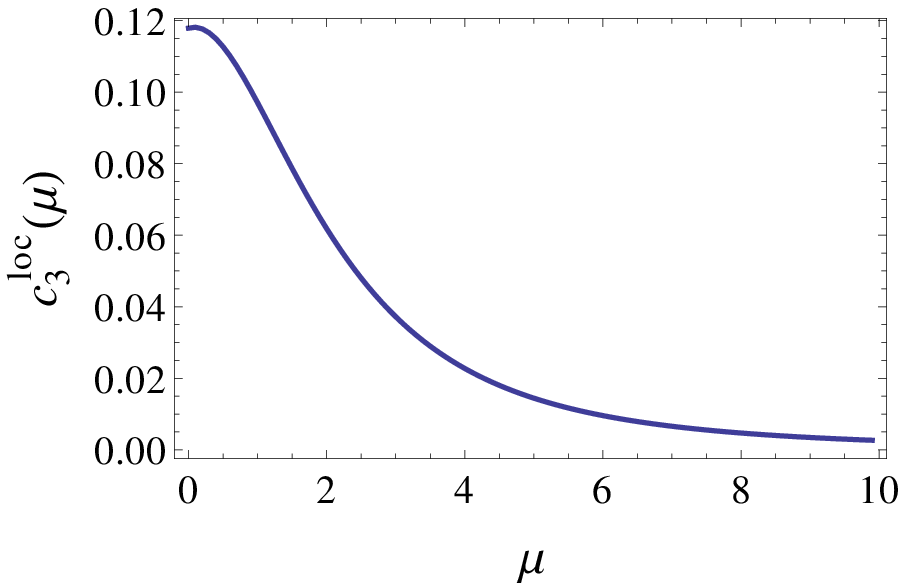}
\includegraphics[scale=0.93]{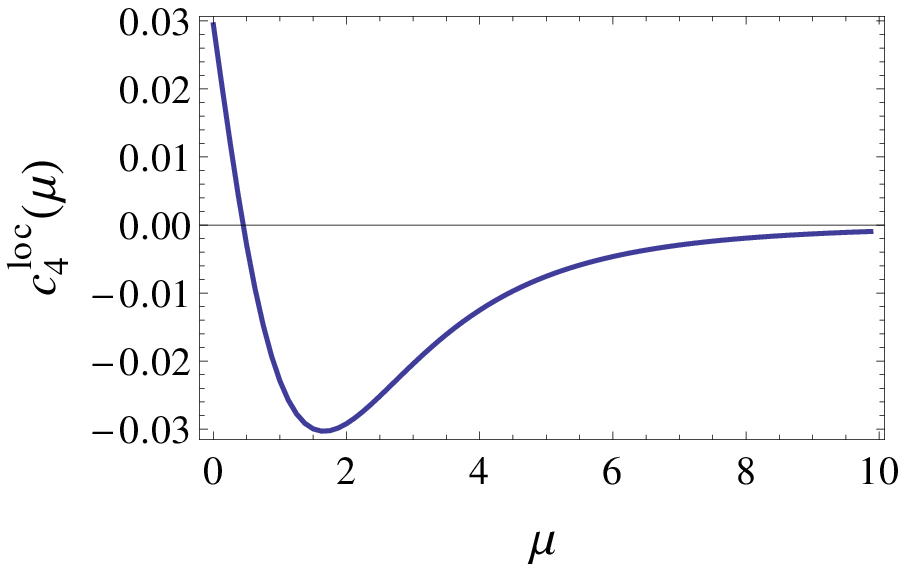}\\
\end{center}
\caption{(Color online)
Coefficients $ c^{\text{loc}}_3(\mu) $ and $ c^{\text{loc}}_4(\mu) $ in the third and the fourth cumulants of the local time of the dry and viscous friction case [see Eq.~(\ref{eh0_ch5})].}
\label{fig_loc_cumulant}
\end{figure*}

To quantify the accuracy of the asymptotic Gaussian distribution we can now check as well for the
size of the higher order cumulants. From Eq.~(\ref{ed_ch5}) we can easily confirm that to leading order
the third and the fourth cumulants satisfy
\begin{equation}
K_3^{\text{loc}}=c_3^{\text{loc}}(\mu) t+ O(1),\qquad K_4^{\text{loc}}=c_4^{\text{loc}}(\mu) t+O(1),
\label{eh0_ch5}
\end{equation}
where both coefficients $ c_3^{\text{loc}}(\mu) $ and $ c_4^{\text{loc}}(\mu) $ are independent of the initial value $ v_0 $.
Intuitively, we expect that we recover the result of the previous subsection [see Eq.~(\ref{dfa_ch5})]
if the dry friction term dominates the viscous friction force. That means that
the coefficients $ c_3^{\text{loc}}(\mu) $ and $ c_4^{\text{loc}}(\mu) $ in Eq.~(\ref{eh0_ch5}) which govern the linear time dependence
of the cumulants decay to zero for large $ \mu $ [see also Eq.~(\ref{dfa_ch5})]. Indeed, numerical evaluation
of the coefficients confirms what one expect intuitively (see Fig.~\ref{fig_loc_cumulant}). It is however remarkable and
somehow counterintuitive that the addition of a viscous force results in large cumulants as compared to the pure dry friction case [see Eq.~\ref{dfa_ch5})]. The linear increase of the cumulants with time indicates that the entirely
Gaussian behavior of large fluctuations of the pure dry friction model, i.e., a strictly quadratic large deviation function,
is modified by non-Gaussian contributions due to viscous damping. In fact,  Fig.~\ref{fig_loc_cumulant} shows
a resonance like phenomenon, where the deviations from Gaussian behavior become maximal at some intermediate value
of $\mu$. The fourth order cumulant becomes maximal in absolute value if
the relative strength of dry and viscous friction attains a certain optimal value.

\section{Occupation time}
\label{sec4}

As we have seen in the previous section, largely closed analytic expressions can be derived for the
distribution function of the local time. To check the robustness of the conclusions against
the change of the observable we now consider the occupation time (\ref{myg}). We
will see that we can again provide a largely analytic account if we still follow the
strategy outlined in the previous section.

\subsection{General drift}

Let us first discuss to which extent we can derive properties of the occupation time distribution
and its moments without specifying the drift. As pointed out in Eq.~(\ref{bh0_ch5}) the construction
of  particular solutions of the inhomogeneous moment hierarchy (\ref{bj0_ch5}) is the key, if we assume
a fundamental solution of the homogeneous equation (\ref{by_ch5}) to be given.

We first observe that the kernel $U(v)=\theta(v)$ vanishes for $v<0$, so that for a particular solution
of the moment hierarchy (\ref{bj0_ch5}), we may assume $\widetilde{M}_n^p(s,v_0)=0$ for $v_0<0$.
Furthermore, the first equation of the hierarchy ($n=1$) can be solved easily to yield
\begin{equation}
\widetilde{M}_1^p(s,v_0)=\left\{
  \begin{array}{lll}
    1/s^2 && \mbox{for } v_0>0,\\
    0 && \mbox{for } v_0<0.
  \end{array}
\right.
\end{equation}
Now to compute the higher order contributions $\widetilde{M}_n^p(s,v_0)$ recursively
we observe that the $m$-th order derivative of the
homogeneous equation (\ref{by_ch5}) with respect to $s$ just resembles the equations of the moment hierarchy, i.e.,
\begin{equation}
\left[\partial_{v_0}^2-\Phi'(v_0)\partial_{v_0}-s \right] \partial_{s}^m \varphi(s,v_0)=
m\partial_{s}^{m-1} \varphi(s,v_0) .
\label{ca_ch5}
\end{equation}
Hence $\widetilde{M}_n^p(s,v_0)$ can be expressed as a linear combination of partial derivatives of the
fundamental solution. Using such an inductive argument and Eq.~(\ref{bh0_ch5}) we finally arrive at the expression for the Laplace transform of the $n$-th moment of the occupation time,
\begin{widetext}
\begin{eqnarray}
\widetilde{M}^{\text{occ}}_n(s,v_0)=
\left\{
\begin{array}{lll}
n!/s^{n+1}+\sum_{m=0}^{n-1}(-1)^m {n\choose m}C^+_{n-m}\partial^m_{s}\varphi(s,v_0) && \mbox{for } v_0>0,\\
C^-_n\varphi(s,v_0) && \mbox{for } v_0<0.
\end{array}
\right.
\label{cb_ch5}
\end{eqnarray}
\end{widetext}
The coefficients  $ C^+_n $ and $ C^-_n $
are determined by the matching condition at $v_0=0$ [see Eqs.~(\ref{bo_ch5}) and (\ref{bq_ch5})].
Thus, we obtain the explicit formulas
\begin{eqnarray}
C^{\pm}_1 = \frac{\partial_{v_0}\varphi(s,0 \mp)}{s^2\varphi(s,0)[\partial_{v_0}\varphi(s,0+)
-\partial_{v_0}\varphi(s,0-)]
},
\label{cc_ch5}
\end{eqnarray}
and for $ n>1 $
\begin{eqnarray}
C^{\pm}_n &=& \Bigg\{\sum_{m=1}^{n-1}(-1)^m {n\choose m} C^+_{n-m}[\partial_{v_0}\varphi(s,0\mp)\partial^m_{s}\varphi(s,0)
\nonumber\\
&&-\varphi(s,0)\partial^m_{s}\partial_{v_0}\varphi(s,0+)]
+\frac{n!\partial_{v_0}\varphi(s,0\mp)}{s^{n+1}}\Bigg\}\nonumber\\
&& \Big / \big\{\varphi(s,0)[\partial_{v_0}\varphi(s,0+)
-\partial_{v_0}\varphi(s,0-)]\big\}.
\label{cd_ch5}
\end{eqnarray}
While such expressions are certainly more involved than those for the local time problem we can still express the entire moment problem in terms of the fundamental solution of the homogeneous
equation (\ref{by_ch5}).

\subsection{Pure dry friction}

\begin{figure*}
\begin{center}
\includegraphics[scale=0.9]{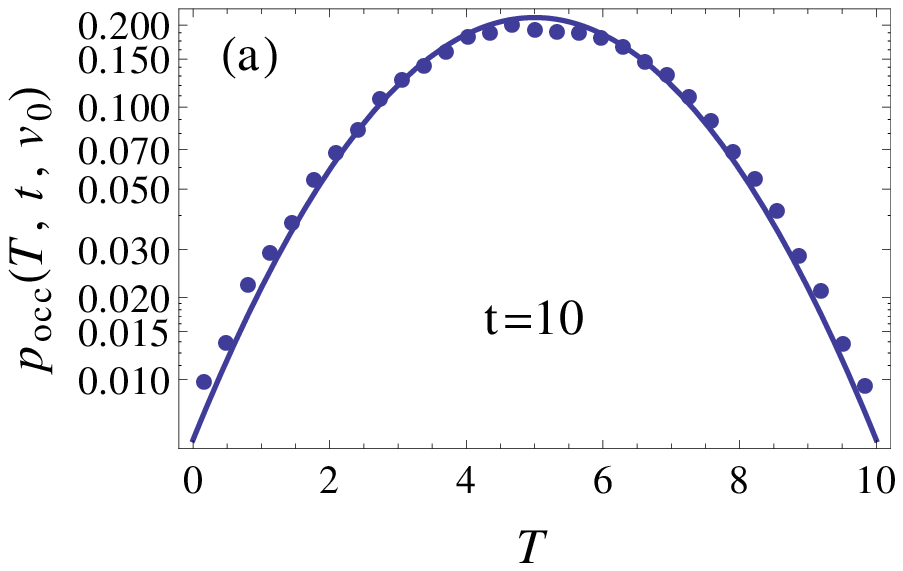}
\includegraphics[scale=0.95]{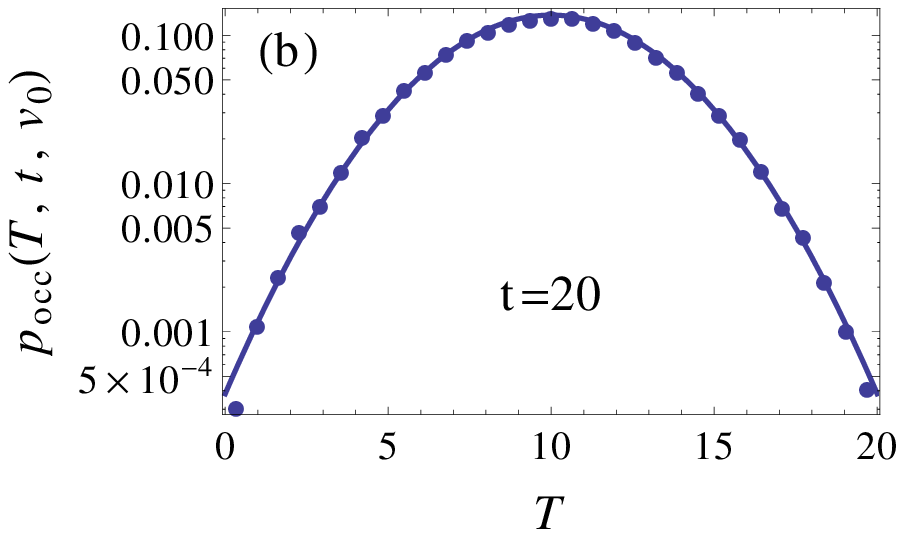}
\end{center}
\caption{(Color online)
Occupation time distribution of the pure dry friction case [see Eqs.~(\ref{ca_ch2}) and (\ref{bh_ch5})] for $ v_0=0 $ and two different times: (a) $ t=10 $ and (b) $ t=20 $. Lines correspond to the leading-order asymptotic distribution (\ref{dk_ch5}),  and points to the Monte Carlo simulation of the corresponding Langevin equation  by using the Euler-Maruyama scheme with time-step $ \Delta t=0.0001 $ and an ensemble of $ 10^6 $ realizations. \yaming{ With the chosen  step-size and ensemble the standard
numerical errors of the Euler-Maruyama scheme (see Ref.~\cite{Gardiner2009StoMeth}) are within the resolution of
the figure, apart from the extreme tails of the distribution where the statistics
becomes poor.}  }
\label{fig_occ_dry_ch5}
\end{figure*}

For the model with only dry friction, i.e., with the the potential (\ref{bh_ch5}), the fundamental solution of the
corresponding homogeneous system is given by Eq.~(\ref{da_ch5}). Hence, it is straightforward to obtain the moments of the occupation time
in the Laplace space by using the formula (\ref{cb_ch5}). The first two moments are given explicitly by
\begin{gather}
\widetilde{M}^{\text{occ}}_{1}(s,v_0) = \frac{1}{2s^2}
\left\{
\begin{array}{lll}
2-e^{-(\sqrt{1+4s}-1)v_0/2} && \mbox{for } v_0>0,\\
e^{(\sqrt{1+4s}-1)v_0/2} && \mbox{for } v_0<0,
\end{array}\right.
 \label{dg_ch5}
\end{gather}
\vspace{-1em}
\begin{widetext}
\begin{equation}\label{dh_ch5}
\widetilde{M}^{\text{occ}}_{2}(s,v_0)  =\frac{1}{4s^3} \left\{
\begin{split}
&8 -
\frac{ 5\sqrt{4 a+1}+4v_0s +1 }{ \sqrt{4s+1} } e^{- \left(\sqrt{4
   s+1}-1\right) v_0/2} && \mbox{for } v_0>0,\\
&\frac{3\sqrt{4 s+1}-1}{\sqrt{4s+1} } e^{\left(\sqrt{4
  s+1}-1\right) v_0/2} && \mbox{for } v_0<0 .
\end{split}
\right.
\end{equation}
\end{widetext}
The analytic structure of these expressions is solely determined by the analytic structure of the fundamental solution and
thus coincides with what we found for the local time, namely a pole at $ s=0 $ and a branch cut for $ s<-1/4 $. Thus the
asymptotic expansion of the inverse Laplace transform results in the long time behavior of the moments,
\begin{gather}
M^{\text{occ}}_{1}(t,v_0)= \frac{t}{2}+\frac{v_0}{2}+O(e^{-(1/4-o)t}),
\label{di_ch5}\\
M^{\text{occ}}_{2}(t,v_0)= \frac{t^2}{4}+\frac{1+v_0}{2}t+\frac{v_0^2-6}{4}+O(e^{-(1/4-o)t}). \label{dj_ch5}
\end{gather}
The variance of the occupation time is easily computed as
\begin{equation}
\text{Var}(T_{\text{occ}})=\frac{t}{2}-\frac{3}{2}+O(e^{-(1/4-o)t}),
\end{equation}
and the limiting distribution (\ref{ec_ch5}) reads
\begin{eqnarray}\label{dk_ch5}
p_{\text{occ}}^{\text{asym}}(T,t,v_0) &=& \frac{1}{\sqrt{\pi (t-3)}}
\exp\left[-\frac{(2T-t-v_0)^2}{4t-12}\right]\nonumber\\
&&+O(e^{-(1/4-o)t}).
\end{eqnarray}
At large time, this limiting distribution matches well with the Monte Carlo simulation of the corresponding Langevin equation by
using the Euler-Maruyama scheme \cite{Kloeden1992NumericalSDE} (see Fig.~\ref{fig_occ_dry_ch5}). \yaming{ As shown in \cite{ChenBaule2013}, the application of this scheme is stable for piecewise-smooth stochastic differential equations and only requires that we choose the integration time-step small enough. }

To estimate whether there are corrections to the leading asymptotic Gaussian behavior of large deviations [see Eq.~(\ref{dk_ch5})] we just work out, as before,
the third and the fourth cumulants
\begin{equation}
K^{\text{occ}}_3=O(e^{-(1/4-o)t}),\quad K^{\text{occ}}_4=-45/4+O(e^{-(1/4-o)t}).
\label{dka_ch5}
\end{equation}
As these quantities show no linear growth with time both become negligible for large deviation
properties and we expect again a purely quadratic large deviation function (see also \cite{Simpson2013opt}). Thus as for the local
time [see Eq.~(\ref{dfa_ch5})] Gaussian behavior prevails.

\subsection{Dry and viscous friction}

\begin{figure*}
\begin{center}
\includegraphics[scale=0.91]{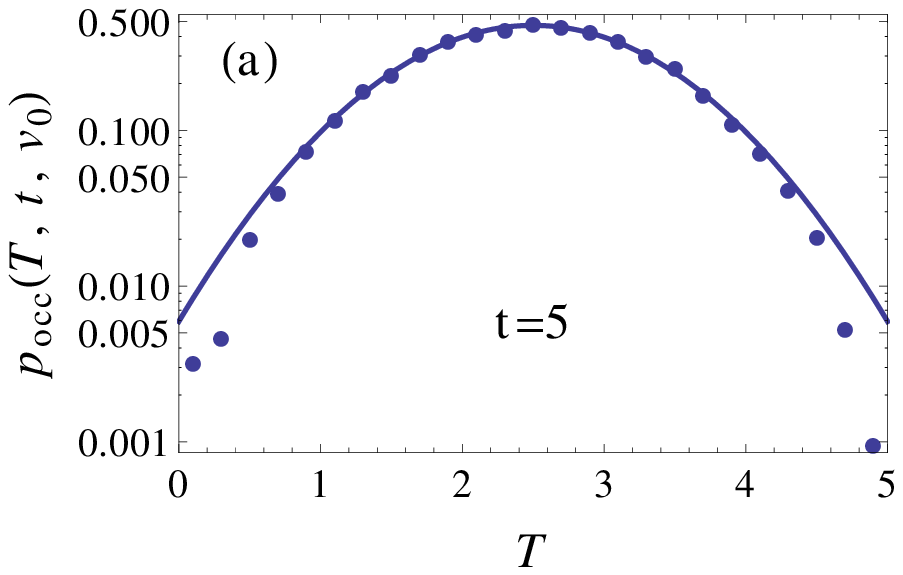}
\includegraphics[scale=0.9]{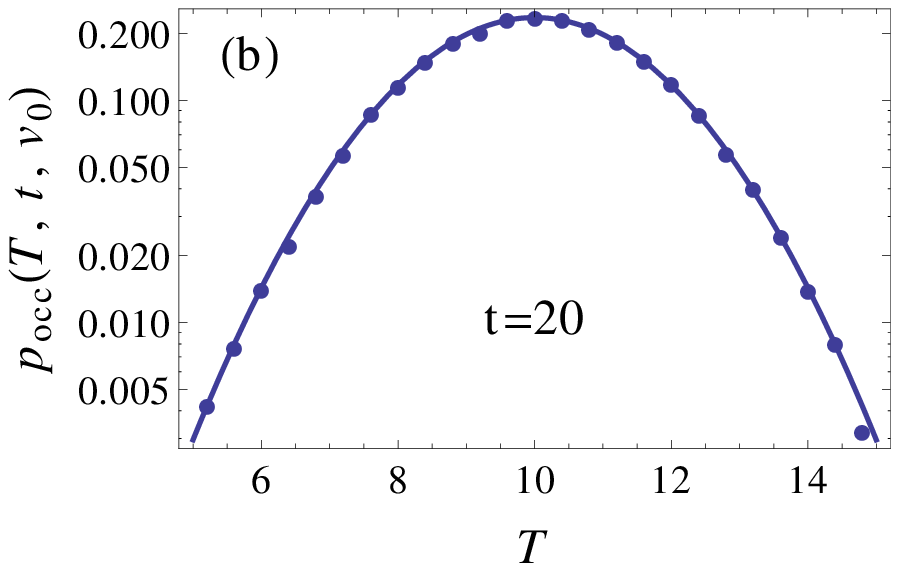}
\end{center}
\caption{(Color online)
Occupation time distribution of the dry and viscous friction case [see Eqs.~(\ref{ca_ch2}) and (\ref{ea_ch5})] for $ v_0=0 $, $ \mu=1 $ and two different times: (a) $ t=5 $ and (b) $ t=20 $. Lines correspond to the leading-order asymptotic distribution (\ref{ec_ch5}), and points to the Monte Carlo simulation of the corresponding Langevin equation by using the Euler-Maruyama scheme with time-step $ \Delta t=0.0001 $ and an ensemble of $ 10^6 $ realizations.}
\label{fig_occ_full_ch5}
\end{figure*}

The inclusion of a viscous force results in the Langevin equation (\ref{ca_ch2}) with the potential (\ref{ea_ch5}).
As for the local time the knowledge of the fundamental solution (\ref{eb_ch5}) is sufficient to evaluate the
moments of the occupation time by Eqs.~(\ref{cb_ch5}), (\ref{cc_ch5}), and (\ref{cd_ch5}). Even though the involvement
of the parabolic cylinder function makes the computations slightly cumbersome it is possible to write
down explicit results for the low order moments. For the Laplace transform of the first two moments we obtain
\begin{widetext}
\begin{eqnarray}
\widetilde{M}^{\text{occ}}_1(s,v_0)=\frac{1}{2 s^2 D_{-s}(\mu)} \left\{
\begin{array}{lll}
  2 D_{-s}(\mu)-e^{v_0^2/4+\mu v_0/2}
   D_{-s}(\mu+v_0) && \mbox{for } v_0>0,\\
  e^{v_0^2/4-\mu v_0/2}
   D_{-s}(\mu-v_0)  && \mbox{for } v_0<0,
\end{array}
\right.
\end{eqnarray}
\vspace{-1em}
\begin{eqnarray}
\widetilde{M}^{\text{occ}}_2(s,v_0)&=&\frac{e^{v_0^2/4 +\mu v_0/2}    D_{-s}(\mu+v_0)}{2  s^3 D_{-s-1}(\mu)
   D^2_{-s}(\mu)}
   \Big\{D_{-s}(\mu) \left[s
   D^{(1,0)}_{-s-1}(\mu)-2
   D_{-s-1}(\mu)\right]
   +D_{-s-1}(\mu) \left[s
   D^{(1,0)}_{-s}(\mu)-D_{-s}(\mu)\right]\Big\}\nonumber\\
   &&+\frac{2}{s^3}
   -\frac{e^{v_0^2/4 +\mu v_0/2}}{s^2D_{-s}(\mu)}D_{-s}^{(1,0)}(\mu+v_0) \quad \mbox{for } v_0>0,
\end{eqnarray}
\vspace{-1em}
\begin{eqnarray}
\widetilde{M}^{\text{occ}}_2(s,v_0) =\frac{e^{v_0^2/4-\mu v_0/2}    D_{-s}(\mu-v_0) }{2  s^3
   D_{-s-1}(\mu) D_{-s}^2 (\mu)}
   \Big\{ D_{-s}(\mu) \Big[s
   D^{(1,0)}_{-s-1}(\mu)
      +D_{-s-1}(\mu)
   \Big]
   -s D_{-s-1}(\mu)
   D^{(1,0)}_{-s}(\mu)\Big\}
   \quad   \mbox{for } v_0<0.
\end{eqnarray}
For the inversion of the Laplace transform in the asymptotic limit of large time we again have to
analyze the singularities of these expressions. Poles appear at $s=0$ and at values where the
parabolic cylinder functions vanish. It is a rather obvious consequence of the properties
of the parabolic cylinder functions (see subsection \ref{subsec_cc}) that all these singularities are poles
on the nonpositive real axis and that the largest negative pole appears at $s_0+1$, where $s_0$ has been
introduced in Eq.~(\ref{ed0_ch5}) (see also Fig.~\ref{fig_s0}). Hence the standard asymptotic expansion
of the inverse Laplace transform results in
\begin{gather}
M^{\text{occ}}_1(t,v_0) = \frac{t}{2}+\frac{\sigma(v_0)}{2}\left(  \frac{D_0^{(1,0)}(\mu+|v_0|)}{D_0(\mu+|v_0|)}-\frac{D_0^{(1,0)}(\mu)}{D_0(\mu)} \right)+O(e^{(s_0+1+o) t}),\label{ei_ch5}\\
M^{\text{occ}}_2(t,v_0)=
\frac{t^2}{4} +\frac{\sigma(v_0)}{2}\left( \frac{D_{0}^{(1,0)}(\mu+|v_0|)}{D_{0}(\mu+|v_0|)}-\frac{D_{0}^{(1,0)}(\mu)}{D_{0}(\mu)} \right) t
+\frac{1}{2}\left( \frac{D_{-1}^{(1,0)}(\mu)}{D_{-1}(\mu)} -\frac{D_{0}^{(1,0)}(\mu)}{D_{0}(\mu)} \right)t+O(1),\label{ej_ch5}
\end{gather}
\end{widetext}
\yaming{where $ s_0+1+o<0 $. } Thus the variance of the occupation time reads
\begin{equation}\label{ek_ch5}
\mathrm{Var}(T_{\text{occ}})= \frac{1}{2}\left( \frac{D_{-1}^{(1,0)}(\mu)}{D_{-1}(\mu)} -\frac{D_{0}^{(1,0)}(\mu)}{D_{0}(\mu)} \right)t+O(1).
\end{equation}
Here we have used, as before, the symbol $D_{\nu}^{(1,0)}$ to denote the derivative of the parabolic cylinder function with respect to its index.
The first moment (\ref{ei_ch5}) and the variance (\ref{ek_ch5}) determine the asymptotic behavior of
the limiting distribution (\ref{ec_ch5}) which, for large values of $t$, matches well with the Monte Carlo simulation of the corresponding
Langevin equation, as shown in Fig.~\ref{fig_occ_full_ch5}.

\begin{figure}
\begin{center}
\includegraphics[scale=0.9]{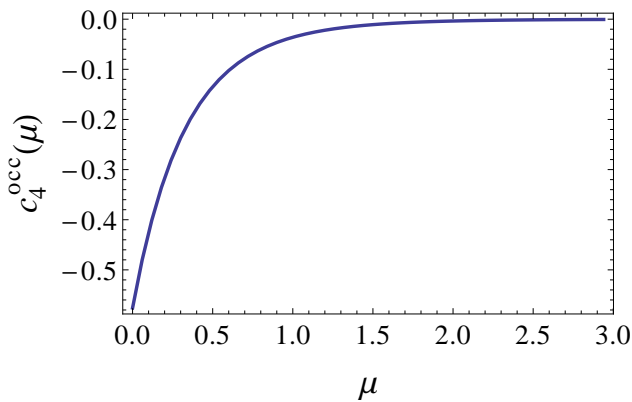}
\end{center}
\caption{(Color online)
Coefficient $ c^{\text{occ}}_4(\mu) $ in the fourth cumulant of the occupation time of dry and viscous friction case [see Eq.~(\ref{ek0_ch5})].}
\label{fig_occ_cumulant}
\end{figure}

In order to quantify deviations from the Gaussian limit we can again
check the values of the third and the fourth cumulants, which are evaluated as
\begin{equation}
K^{\text{occ}}_3=O(1),\qquad K^{\text{occ}}_4=c_4^{\text{occ}}(\mu) t+O(1),
\label{ek0_ch5}
\end{equation}
where $ c_4^{\text{occ}}(\mu) $ is independent of the initial value $ v_0 $. While it is quite straightforward to express
such a coefficient in terms of parabolic cylinder functions we just focus here on a graphical discussion of the result.
As mentioned in the previous section we expect the cumulants to approach the dry friction case [see
Eq.~(\ref{dka_ch5})] in the limit when $\mu$ tends to infinity. Indeed the coefficient $ c^{\text{occ}}_4(\mu) $  in
Eq.~(\ref{ek0_ch5}) decays to zero, as shown in Fig.~\ref{fig_occ_cumulant}. Furthermore, the occurrence of a contribution which is linear in time
means that deviations from a purely quadratic large deviation function appear due to the inclusion of
the viscous force, as it was the case for the local time. Hence such a feature does not seem to be
related to the particular functional that is considered.

\section{Displacement}
\label{sec5}

Finally we investigate the distribution of the displacement of the object, which as well can be expressed
in terms of the functional (\ref{mya}) with the choice $U(v)=v$.
In this case, particular solutions $ \widetilde{M}_n^p(s,v_0) $ of the corresponding inhomogeneous ODE
(\ref{bd_ch5}) are not easy to write down in general. So a general formula like Eq.~(\ref{bg_ch5}) or (\ref{cb_ch5})
is not available. Fortunately, for the dry friction model (\ref{ad_ch2}) we will see in the following subsections
that particular solutions can be constructed rather straightforwardly for the pure dry friction case and the full model,
respectively.

\subsection{Pure dry friction}

Let us consider Eq.~(\ref{bd_ch5}) with the potential (\ref{bh_ch5}) and $U(v)=v$ for $n=1$. On either of the domain
$v_0>0$ or $v_0<0$, it is obvious that a particular polynomial solution is given by
\begin{equation}\label{dl_ch5}
\widetilde{M}_1^p(s,v_0)=\frac{s v_0-\sigma(v_0)}{s^3}.
\end{equation}
Hence, using Eq.~(\ref{da_ch5}) for the homogeneous solution we obtain from Eq.~(\ref{bh0_ch5})
the expression for the first moment. The two constants are then determined by the two matching conditions
(\ref{bo_ch5}) and (\ref{bq_ch5}) so that we arrive at
\begin{eqnarray}
\hspace{-2em}\widetilde{M}^{\text{dis}}_1(s,v_0)&=&
\frac{s v_0-\sigma(v_0)}{s^3} +\frac{\sigma(v_0)}{s^3} \varphi(s,v_0)\nonumber\\
&=& \frac{s v_0-\sigma(v_0)}{s^3} +\frac{\sigma(v_0)}{s^3}e^{ (1-\sqrt{1+4s})|v_0|/2 }.
\label{dm_ch5}
\end{eqnarray}
For the moment of the next order we have to determine a particular solution of the inhomogeneous equation
(\ref{bd_ch5}) for $n=2$, i.e., where the right hand side is essentially given by Eq.~(\ref{dm_ch5}). While that
is certainly a straightforward task we can uncover the underlying algebraic structure by reducing
all the manipulations to the properties of the fundamental solution (\ref{da_ch5}).
Since
\begin{eqnarray}
|v_0| \varphi(s,v_0)&=&|v_0|e^{ (1-\sqrt{1+4s})|v_0|/2 }\nonumber\\
&=&-\sqrt{4s+1}\partial_s e^{ (1-\sqrt{1+4s})|v_0|/2 }\nonumber\\
&=&
-\sqrt{4s+1}\partial_s \varphi(s,v_0) ,
\label{dma_ch5}
\end{eqnarray}
we can use the property (\ref{ca_ch5}) for $m=1$ to construct one particular solution of Eq.~(\ref{bd_ch5})
for $ n=2 $ as
\begin{eqnarray}
\widetilde{M}_2^p(s,v_0)&=&\frac{\sqrt{1+4s}v_0^2+2|v_0|}{s^3(1+4s)}e^{(1-\sqrt{1+4s})|v_0|/2}\nonumber\\
&&+\frac{(2s|v_0|-3)^2+8s+3}{2s^5} .
\label{dn_ch5}
\end{eqnarray}
The first term in this expression
is caused by the second term in the inhomogeneity (\ref{dm_ch5}), while the second term is a polynomial part due to Eq.~(\ref{dl_ch5}). For this contribution we have
used the property (\ref{dma_ch5}). Finally the second moment is determined by Eq.~(\ref{bh0_ch5}) and
the matching conditions (\ref{bo_ch5}) and (\ref{bq_ch5}), resulting in
\begin{eqnarray}
\widetilde{M}^{\text{dis}}_2(s,v_0)&=&
-\frac{(11s+3)\left(\sqrt{4 s+1}+1\right)}{s^5 (4 s+1)}e^{(1-\sqrt{1+4s})|v_0|/2}\nonumber\\
&&+\widetilde{M}_2^p(s,v_0).
\label{do_ch5}
\end{eqnarray}
Higher order moments can now be constructed easily by using these algebraic steps repeatedly.

\begin{figure*}
\begin{center}
\includegraphics[scale=0.95]{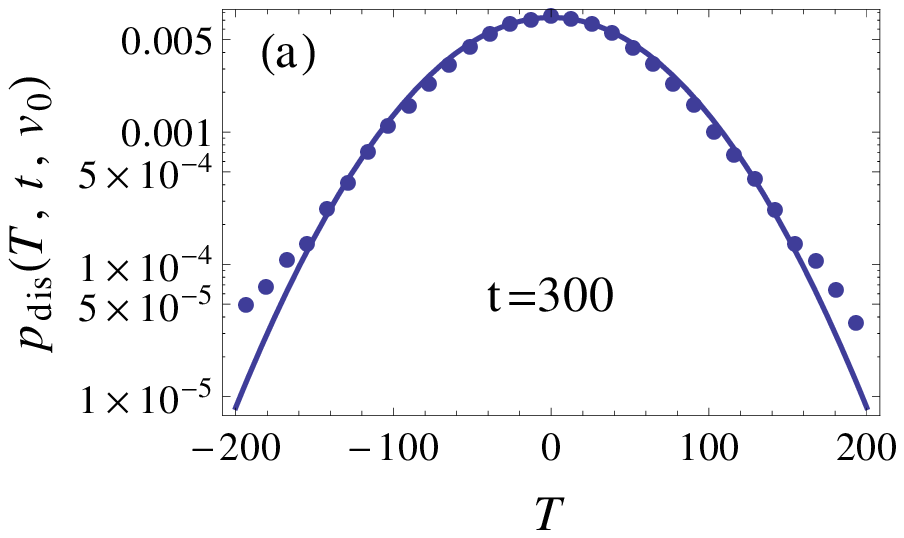}
\includegraphics[scale=0.9]{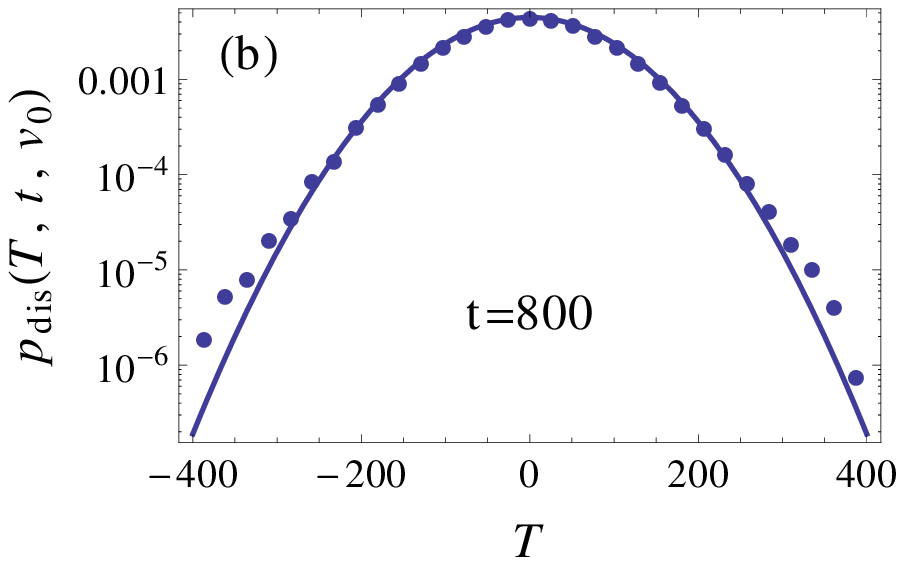}
\end{center}
\caption{(Color online)
Displacement distribution of the pure dry friction case [see Eqs.~(\ref{ca_ch2}) and (\ref{bh_ch5})] for $ v_0=0 $ and two different times: (a) $ t=300 $ and (b) $ t=800 $. Lines correspond to the leading-order asymptotic distribution (\ref{du_ch5}),  and points to the Monte Carlo simulation of the corresponding Langevin equation by using the Euler-Maruyama scheme with time-step $ \Delta t=0.001 $ and an ensemble of $ 10^6 $ realizations.}
\label{fig_dis_dry_ch5}
\end{figure*}

As for the analytic structure, the approach described above clearly shows that
the moments inherit their analytic structure from the homogeneous solution. Thus the first two moments
have a pole at $ s=0 $ and a branch
cut for $ s<-1/4 $. The usual asymptotic expansion for the inverse Laplace transform results in
\begin{eqnarray}
M^{\text{dis}}_1(t,v_0) = v_0(1+|v_0|/2)+O(e^{-(1/4-o)t}),\label{ds_ch5}
\end{eqnarray}
\begin{eqnarray}
M^{\text{dis}}_2(t,v_0) &=& 10t+v_0^4/4+5|v_0|^3/3+5v_0^2-54\nonumber\\
&&+O(e^{-(1/4-o)t}).
\label{dt_ch5}
\end{eqnarray}
Hence we obtain for the variance of the displacement
\begin{eqnarray}
\text{Var}(T_{\text{dis}})&=&10t+2|v_0|^3/3+4v_0^2-54+O(e^{-(1/4-o) t})\nonumber\\
&=&\sigma^2(t) +O(e^{-(1/4-o) t}),
\end{eqnarray}
and the limiting distribution (\ref{ec_ch5}) reads
\begin{eqnarray}
p_{\text{dis}}^{\text{asym}}(T,t,v_0)\!\!\!&=& \!\!\!
\frac{1}{\sqrt{2\pi \sigma^2(t) }}\exp\left[-\frac{(T-v_0-v_0|v_0|/2)^2}{2 \sigma^2(t)}\right]\nonumber\\
&&+O(e^{-(1/4-o)t}).
\label{du_ch5}
\end{eqnarray}
Around the mean of the displacement, this asymptotic expression matches well with the Monte Carlo simulation of the corresponding Langevin equation, as shown in Fig.~\ref{fig_dis_dry_ch5}.

As already mentioned, the recursive procedure described at the beginning of this subsection
allows also for the computation of the higher order moments, and thus for the evaluation
of the higher order cumulants. However, as the formulas are too cumbersome a software for symbolic calculation,
such as Mathematica, is useful to handle the expressions. We finally obtain for the third and the fourth cumulants
in the long time limit,
\begin{equation}
K^{\text{dis}}_3=\left\{\begin{array}{lll}
O(1) && \mbox{for } v_0\neq 0, \\
0 && \mbox{for } v_0=0,
\end{array}
\right. \qquad K^{\text{dis}}_4=21120 t+O(1).
\label{dua_ch5}
\end{equation}
In this case the higher order cumulants show a linear time dependence already in the pure dry friction model. Hence the corresponding large deviation function
does not show a plain quadratic form any more and the strictly Gaussian behavior does not show up.
Such a feature is as well clearly visible in Fig. \ref{fig_dis_dry_ch5}.
It would be worth to explore whether this difference to the previous cases is related to a functional with an
unbounded kernel.

\subsection{Dry and viscous friction}

\begin{figure*}
\begin{center}
\includegraphics[scale=0.9]{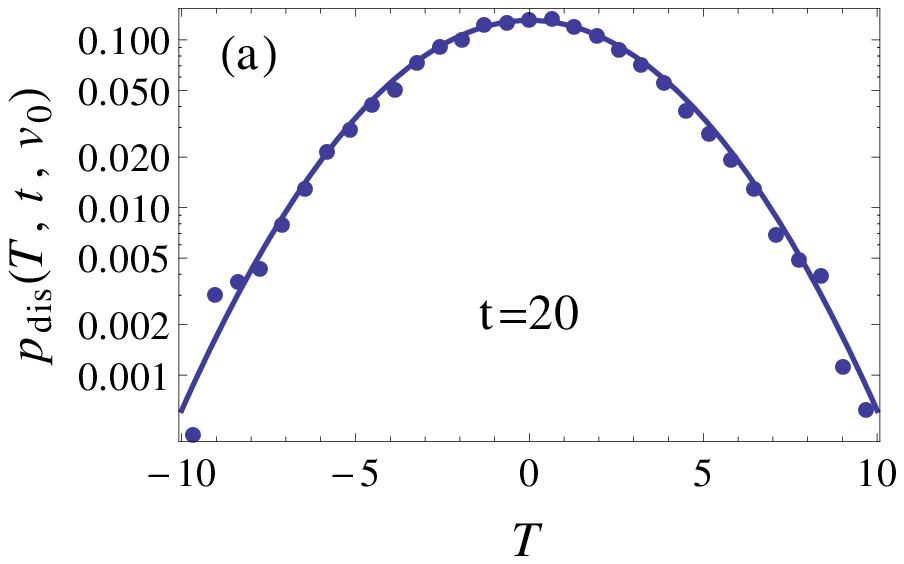}
\includegraphics[scale=0.94]{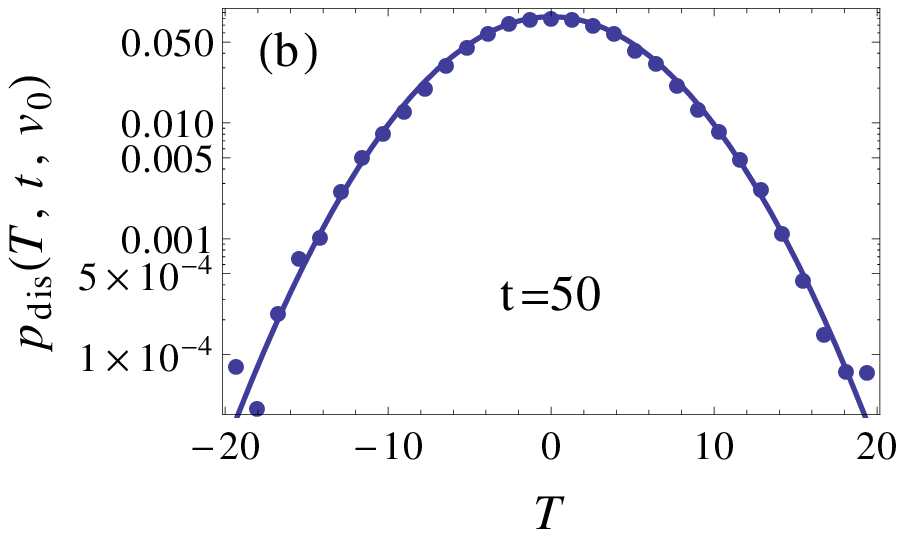}
\end{center}
\caption{(Color online) Displacement distribution of the dry and viscous friction case [see Eqs.~(\ref{ca_ch2}) and (\ref{ea_ch5})] for $ v_0=0 $, $ \mu=1 $ and two different times: (a) $ t=20 $ and (b) $ t=50 $. Lines correspond to the leading-order asymptotic distribution of Eq.~(\ref{ec_ch5}), and points to the Monte Carlo simulation of the corresponding Langevin equation by using the Euler-Maruyama scheme with time-step $ \Delta t=0.001 $ and an ensemble of $ 10^6 $ realizations.}
\label{fig_dis_full_ch5}
\end{figure*}

We now apply the procedure described in the previous subsection to analyze the distribution of the
displacement in a model with dry and viscous friction. To begin with
one particular polynomial solution of Eq.~(\ref{bd_ch5}) with $ U(v)=v $,
potential (\ref{ea_ch5}), and $ n=1 $ is given by
\begin{equation}\label{el_ch5}
\widetilde{M}_1^{p}(s,v_0)=
          \frac{s v_0-\mu \sigma(v_0)}{s^2 (s +1)}.
\end{equation}
Hence using the expression for the general solution (\ref{bh0_ch5}) and the matching conditions
at the discontinuity to determine the constants of integration we arrive at [see Eq.~(\ref{bh0_ch5})]
\begin{widetext}
\begin{equation}\label{em_ch5}
\widetilde{M}^{\text{dis}}_1(s,v_0)=
\frac{s v_0-\mu \sigma(v_0)}{s^2 (s +1)}+\mu\sigma(v_0)\frac{
\varphi(s,v_0)}{s^2 (s+1) \varphi(s,0)}=
\frac{s v_0-\mu \sigma(v_0)}{s^2 (s +1)}+\mu\sigma(v_0)\frac{   e^{(\mu+|v_0|)^2/4 }
D_{-s}(\mu+|v_0|)  }{s^2 (s+1) e^{\mu^2/4}D_{-s}(\mu)} .
\end{equation}
For the moment of the next order we need to integrate the inhomogeneous equation (\ref{bd_ch5}) where
the right hand side is essentially given by the first order moment (\ref{em_ch5}). As before we can
make use of algebraic properties of the fundamental solution $\varphi(s,v_0)$ to construct $\widetilde{M}^p_2(s,v_0)$.
Using the recurrence relation of the parabolic cylinder function  (see, e.g., \cite{CuytPetersen2008}), we have
\begin{eqnarray}
\hspace{-2em}|v_0|\varphi(s,v_0) \!\!&=&\!\!
|v_0|e^{(\mu+|v_0|)^2/4}D_{-s}(\mu+|v_0|)= e^{(\mu+|v_0|)^2/4}\big[
D_{-s+1}(\mu+|v_0|)
-s D_{-s-1}(\mu+|v_0|) -\mu D_{-s}(\mu+|v_0|)
 \big]\nonumber \\
\!\!&=&\!\! \varphi(s-1,v_0) -s \varphi(s+1,v_0) -\mu \varphi(s,v_0),
 \label{ema_ch5}
\end{eqnarray}
so that the inhomogeneous part contains a polynomial in $v_0$ and the fundamental solutions
$\varphi(s,v_0)$ and $\varphi(s\pm 1,v_0)$. If we now employ a suitable linear
combination of Eq.~(\ref{ca_ch5}) and the obvious identity [see Eq.~(\ref{by_ch5})]
\begin{equation}
\left[\partial_{v_0}^2-\Phi'(v_0)\partial_{v_0}-s \right] \varphi(s+m,v_0)= m \varphi(s+m,v_0),
\label{myh}
\end{equation}
then a particular solution can be written in terms of the fundamental solution and a polynomial,
namely
\begin{eqnarray}
\widetilde{M}_2^{p}(s,v_0) &=&
\frac{2\mu e^{(\mu+|v_0|)^2/4}}{s^2(s +1)e^{\mu^2/4}D_{-s}(\mu)} \big[D_{-s+1}(\mu +|v_0|)+ s D_{-s-1}(\mu +|v_0|)
+\mu\partial_s D_{-s}(\mu +|v_0|) \big]\nonumber\\
&&+\frac{2v_0^2}{s (s+1) (s+2)}-\frac{2\mu (3s +2) |v_0|}{s^2 (s+1)^2 (s+2)}
+
\frac{4(s+\mu^2)(s+1)+2\mu^2s}{s^3(s+1)^2(s+2)},
\label{en_ch5}
\end{eqnarray}
where the first term is caused by the second term in Eq.~(\ref{em_ch5}) and the last three terms are a polynomial part due to Eq.~(\ref{el_ch5}).
As before, using the matching conditions we then arrive at
\begin{eqnarray}
\widetilde{M}^{\text{dis}}_2(s,v_0)&=&\widetilde{M}^p_2(s,v_0)-\frac{2\mu e^{(|v_0|+\mu)^2/4}D_{-s}(|v_0|+\mu)}{s^3(s+1)^2(s+2) e^{\mu^2/4}D_{-s}(\mu)D_{-s-1}(\mu)}
\Big\{
\mu (s+1)(s+2)\left[D_{-s-1}(\mu)-s D^{(1,0)}_{-s-1}(\mu)\right]\nonumber\\
&&+s(s^2+2s+2)D_{-s}(\mu)+s(s+1)^2(s+2)D_{-s-2}(\mu)
\Big\}. \label{myi}
\end{eqnarray}
\end{widetext}
Higher order moments can be obtained by this algebraic method [see Eqs.~(\ref{ca_ch5}), (\ref{ema_ch5}) and (\ref{myh})]
in a straightforward way.
By expansion, we can check that the points $ s=-1 $ and $ s=-2 $ in Eqs.~(\ref{em_ch5}) and (\ref{myi})
are removable singularities. As for the previous functionals the singularities of the first two moments lie on the nonpositive real axis, and the subleading pole is given by
$ s_0+1 $, where $ s_0 $ is defined in Eq.~(\ref{ed0_ch5}). Hence, the standard asymptotic Laplace inversion yields
for the moments in the time domain
\begin{widetext}
\begin{gather}
M^{\text{dis}}_1(t,v_0)= v_0+\mu\sigma(v_0)\left(\frac{D_0^{(1,0)}(\mu)}{D_0(\mu)}-\frac{D_0^{(1,0)}(\mu+|v_0|)}{D_0(\mu+|v_0|)}\right)+O(e^{(s_0+1+o) t}),\label{eo_ch5}\\
M^{\text{dis}}_2(t,v_0) =
2\left[
1+\mu^2\left(\frac{D_{-1}^{(1,0)}(\mu)  }{D_{-1}(\mu)}-\frac{D_0^{(1,0)}(\mu)}{D_0(\mu)}\right)-2\mu\frac{D_{-2}(\mu)}{D_{-1}(\mu)}
\right]t+O(1) \label{ep_ch5},
\end{gather}
\yaming{where $ s_0+1+o<0 $. }
Therefore the variance reads
\begin{equation}\label{eq_ch5}
\mathrm{Var}(T_{\text{dis}})=
2\left[1+\mu^2\left(\frac{D_{-1}^{(1,0)}(\mu)  }{D_{-1}(\mu)}-\frac{D_0^{(1,0)}(\mu)}{D_0(\mu)}\right)-2\mu\frac{D_{-2}(\mu)}{D_{-1}(\mu)}
\right]t+O(1).
\end{equation}
\end{widetext}
As shown in Fig.~\ref{fig_dis_full_ch5}, the corresponding limiting distribution (\ref{ec_ch5}) in leading order  matches well with the Monte Carlo simulation of the corresponding Langevin equation.

\begin{figure}
\begin{center}
\includegraphics[scale=0.9]{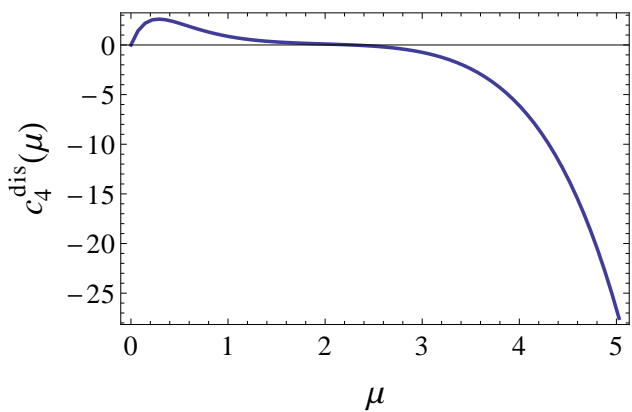}
\end{center}
\caption{(Color online)
Coefficient $ c_4^{\text{dis}}(\mu) $ in the fourth cumulant of the displacement of dry and viscous friction case [see Eq.~(\ref{eq1})].}
\label{fig_dis_cumulant}
\end{figure}

As mentioned above, the higher moments can also be obtained analytically if one uses the properties
(\ref{ca_ch5}), (\ref{ema_ch5}) and (\ref{myh}). But we resort to computer algebra packages to perform the corresponding
tedious formal calculations.
Finally, we confirm that the third and the fourth cumulants satisfy
\begin{equation}
K^{\text{dis}}_3=\left\{\begin{array}{lll}
O(1) && \mbox{for } v_0\neq 0,\\
0 && \mbox{for } v_0=0,
\end{array}
\right. \qquad K^{\text{dis}}_4=c_4^{\text{dis}}(\mu)t+O(1),
\label{eq1}
\end{equation}
which are consistent with those of the pure dry friction case [see Eq.~(\ref{dua_ch5})].
Here the coefficient $ c_4^{\text{dis}}(\mu) $, which is shown in Fig.~\ref{fig_dis_cumulant}
as a function of $ \mu $, does not depend on the initial value $ v_0 $.
From our analytic results we recover that in the case $ \mu=0 $ the third and the fourth cumulants vanish
as the dry and viscous friction model reduces to the Ornstein-Uhlenbeck process, for which the displacement
distribution is strictly Gaussian. In addition, $ c_4^{\text{dis}}(\mu) $ almost vanishes in the region
$1\leqslant \mu \leqslant 3$ (see Fig.~\ref{fig_dis_cumulant}) so that deviations from Gaussian behavior are hardly noticeable (see Fig. ~\ref{fig_dis_full_ch5}).

\section{Conclusion}
\label{sec6}

Large deviation properties of functionals, here integrals over a stochastic process,
have been analyzed for piecewise-smooth stochastic models. We have, in particular, considered the
Brownian motion with dry and viscous friction as an illustrative example for our approach,
where we have derived explicit results for three particular choices of a functional, the local time
which measures the crossing of the discontinuity, the occupation time measuring the
direction of movement of an object, and the spatial displacement.

To cope with general underlying stochastic models, we have extended the backward Fokker-Planck
technique developed in \cite{MajumdarComtet2002OTD} to provide analytic solutions
for a general support functional. For a generic Langevin equation, the hierarchy of
differential equations for the  moments of functionals have been derived and solutions have
been provided for the local time and the occupation time in terms of the solution of the
corresponding homogeneous equation.

Our study of a piecewise-linear model, i.e., the Brownian motion with pure dry friction serves as a simple
case study. For all three functionals the Laplace transform of the moments can be calculated analytically.
From such results asymptotic properties in the time domain are easily extracted. In principle one could
even invert the transform analytically as in this case the propagator of the corresponding
Fokker-Planck equation is known in closed analytic form (see, e.g., \cite{TouchetteStraeten2010Brownian}). The results confirm that the distribution
of the functionals become strictly Gaussian in the asymptotic limit, for the local and for the occupation time, as
also clearly demonstrated by comparing the analytic results with Monte Carlo simulations. Higher order
cumulants contribute for the displacement in the long time limit, so that a strictly quadratic large deviation
function gets modified by these terms.

\begin{figure*}
\centering
\includegraphics[scale=0.9]{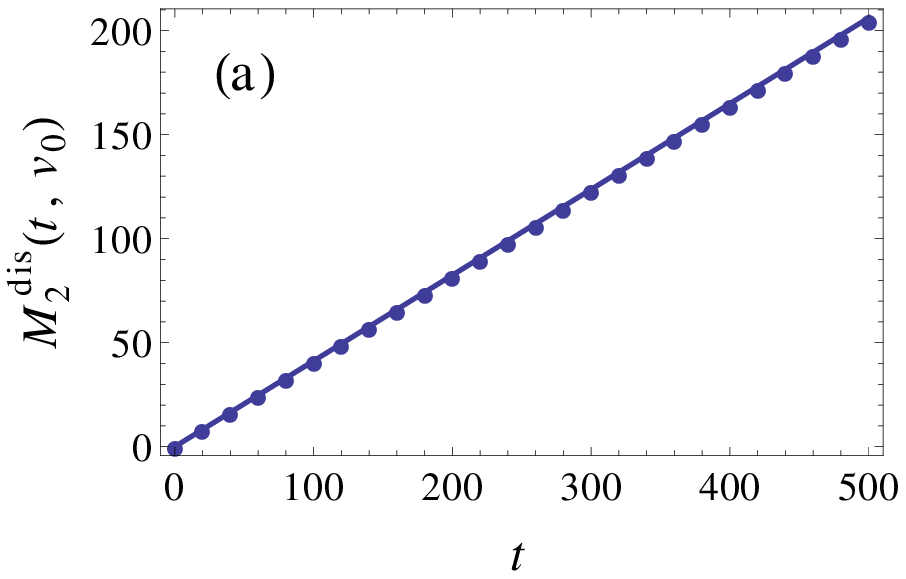}
\includegraphics[scale=0.9]{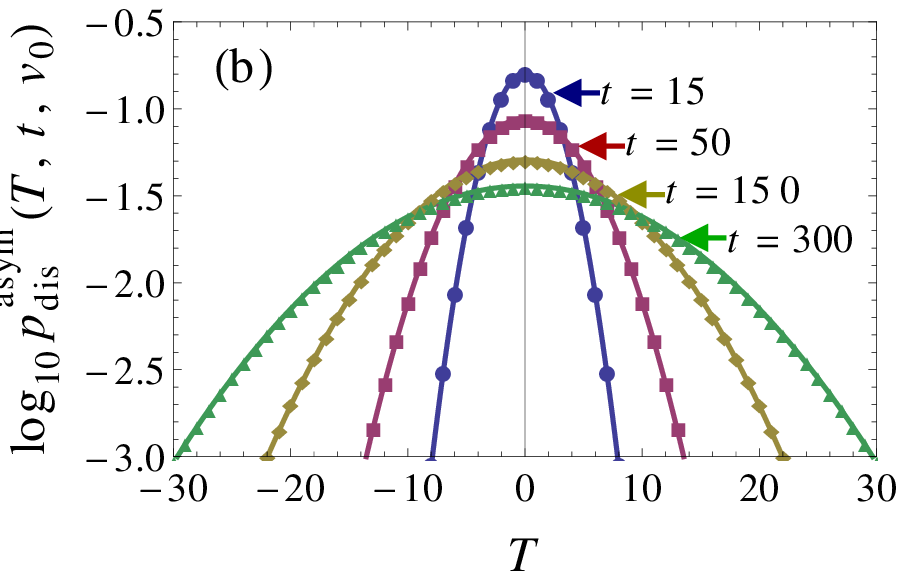}
\caption{(Color online) Analytic results of the dry and viscous friction case in the original scaling system (\ref{ad_ch2}) for $ \gamma=1 $ and $ v_0=0 $. (a) The second moment in leading order obtained from Eq.~(\ref{ep_ch5}) via the rescaling (\ref{cona}); (b) the limiting displacement distribution in leading order obtained from Eqs.~(\ref{ec_ch5}), (\ref{eo_ch5}) and (\ref{eq_ch5}) via the rescaling (\ref{cona}). Lines correspond to the results of $ \mu=6 $ and $ D=5.4 $, and points to $ \mu=1.1 $ and $ D=1 $. These two sets of arguments are properly chosen according to a
fluctuation dissipation relation presented in \cite{MenzelGoldenfeld2011displacement}, ensuring that the numerical results coincide with each other. }\label{Fig.3}
\end{figure*}

The inclusion of an additional viscous force introduces, somehow counterintuitively, non-Gaussian
behavior for the large deviation properties of the local time and the occupation time.
In addition, our analytic results confirm the numerical analysis provided in
\cite{MenzelGoldenfeld2011displacement}  by solving the corresponding Fokker-Planck equation numerically.
For that purpose we need to specialize our general
expressions to the case $ v_0=0 $ and translate our expressions, given in nondimensional units,
to the original scale (\ref{ad_ch2}) via (see \cite{TouchetteStraeten2010Brownian})
\begin{equation}
\mu\rightarrow \mu/(\gamma D)^{1/2}, \quad M_2(t,v_0)\rightarrow DM_2(\gamma t,(\gamma /D)^{1/2}v_0)/\gamma.
\label{cona}
\end{equation}
In Fig.~\ref{Fig.3} it is clearly visible that our analytic expressions confirm
the numerical results presented in \cite{MenzelGoldenfeld2011displacement} (see Figs.~5 and 9 therein)
for the second moment and the distribution. As we have been able to work out higher order cumulants
as well we can even quantify deviations from the predominantly
observed Gaussian behavior.

The analytic method and the results obtained here allow for a rather detailed study on how fluctuations and
related large deviation properties are affected by an underlying piecewise-smooth stochastic dynamics. Even
a simple case, e.g., the extension of the model considered here by
an additional constant bias (see, e.g., \cite{TouchetteStraeten2010Brownian}) seems to be
promising to analyze in detail, as for instance with regards to stick-slip transitions
on the one hand, and even
to benchmark theoretical investigations with real experiments \cite{ChaudhuryMettu2008Brownian,GoohpattaderMettu2009Experimental,
GoohpattaderChaudhury2010Diffusive,GoohpattaderChaudhury2012} on the other.

\begin{acknowledgments}
Y.C. was supported by the Chinese Scholarship Council. We thank Hugo Touchette for useful discussions at the beginning of this research.
\end{acknowledgments}

\end{document}